\documentclass[useAMS,times]{mn2e}

\title[Pseudo-$C_\ell$ estimators and covariances]{CMB temperature and
polarization pseudo-$C_\ell$ estimators and covariances}
\author[M.L. Brown et al.]{M.L. Brown\thanks{E-mail: mlb@roe.ac.uk
(MLB); pgc@roe.ac.uk (PGC); ant@roe.ac.uk (ANT)}, 
P.G. Castro$^{\star}$ \& A.N. Taylor$^{\star}$\\
Institute for Astronomy, University of Edinburgh, Royal Observatory,
Blackford Hill, Edinburgh, EH9 3HJ, UK} 

\usepackage{times}
\def\bib{\parskip=0pt\par\noindent\hangindent\parindent
    \parskip =2ex plus .5ex minus .1ex}

\newcommand{\be}{\begin{equation}}
\newcommand{\ee}{\end{equation}}
\newcommand{\ba}{\begin{eqnarray}}
\newcommand{\ea}{\end{eqnarray}}

\newcommand{\nn}{\nonumber \\}

\newcommand{\rgl}{\rangle} 
\newcommand{\lgl}{\langle}
 
\newcommand{\ld}{{\ell '}}
\newcommand{\ldd}{{\ell ''}} 
\newcommand{\lm}{{\ell m}}
\newcommand{\lmd}{{\ell' m'}} 
\newcommand{\lmdd}{{\ell'' m''}}
\newcommand{\llmm}{{\ell \ell' m m'}} 
\newcommand{\lld}{{\ell \ell'}}
\newcommand{\mmd}{{m m'}} 
\newcommand{\lone}{\ell_1}
\newcommand{\ltwo}{\ell_2}
\newcommand{\mone}{m_1}
\newcommand{\mtwo}{m_2}
\newcommand{\Ylm}{Y_\lm}
\newcommand{\sYlm}{\,_sY_\lm}
\newcommand{\sYlmd}{\,_sY_\lmd} 

\newcommand{\tYlm}{\,_2\!Y_\lm} 
\newcommand{\mtYlm}{\,_{-2}\!Y_\lm}

\pagerange{\pageref{firstpage}--\pageref{lastpage}}
\pubyear{2004}

\label{firstpage}

\begin{document}
%\onecolumn

\maketitle

\begin{abstract}
We develop the pseudo-$C_\ell$ method for reconstructing the Cosmic
Microwave Background (CMB) temperature and polarization auto- and
cross-power spectra, and estimate the pseudo-$C_\ell$ covariance matrix
for a realistic experiment on the cut sky. We calculate the 
full coupling equations for all six possible CMB power spectra, relating
the observed pseudo-$C_\ell$'s to the underlying all-sky $C_\ell$'s, and
test the reconstruction on both full-sky and cut-sky simulated CMB
data sets. In particular we consider the reconstruction of the $C_\ell$ 
from upcoming ground-based polarization experiments covering areas of
a few hundred deg$^2$ and find that the method is fast, unbiased and
performs well over a wide range of multipoles from  $\ell=2$ to $\ell=2500$.
We then calculate the full covariance matrix between the modes of the
pseudo-temperature and polarization power spectra, assuming that the
underlying CMB fields are Gaussian randomly distributed. The complexity of the
covariance matrix prohibits its rapid calculation, required for
parameter estimation. Hence we present an approximation for the
covariance matrix in terms of convolutions of the underlying power
spectra. The coupling matrices in these expressions can be estimated
by fitting to numerical simulations, circumventing direct and slow
calculation, and further, inaccurate analytic approximations. 
We show that these coupling matrices are mostly independent of
cosmology, and that the full covariance matrix for all six pseudo-$C_\ell$
power spectra can be quickly and accurately calculated for any given 
cosmological model using this method. We compare these semi-analytic 
covariance matrices against simulations and find good agreement, the
accuracy of which depends mainly on survey area and the range of
cosmological parameters considered.

\end{abstract}


%%%%%%%%%%%%%%%%%%%%%%%%%%%%%%%%%%%%%%%%%%%%%%%%%%%%%%%%%%%%%%%%%%%%%%
\vspace{-5mm}
\section{Introduction}
The study of the CMB has undergone a revolution in the
past few years, culminating in the first-year Wilkinson Microwave
Anisotropy Probe (WMAP) observations (Bennet et al. 2003). 
The WMAP results along with those from large scale galaxy surveys such
as the 2-degree Field Galaxy Redshift Survey (2dFGRS; 
Percival et al. 2001) and the Sloan Digital Sky Survey (SDSS; 
Tegmark et al. 2004) have revolutionised our understanding of the
Universe, yielding measurements of cosmological
parameters to unprecedented accuracy and providing robust tests of the
standard cosmological model. Furthermore, the pace of progress in
observations of the CMB continues unabated. Following on from the
exquisite WMAP observations of the CMB temperature field, there is a 
number of ground-based and balloon-borne experiments designed to
measure the polarization of the CMB radiation. Indeed the even-parity
E-mode polarization signal has already been detected, initially by the
Degree Angular Scale Interferometer (DASI) experiment (Kovac et
al. 2002). The first-year WMAP observations then yielded a 
detection of the cross correlation between the temperature and the 
E-mode polarization (Kogut et al. 2003) and very recently, the Cosmic
Background Imager (CBI) has detected both the E-mode signal and its
cross correlation with the temperature field (Readhead et al. 2004). 

Over the next few years, one of the central goals of CMB research will
be to tie down the polarization power spectra and take advantage of
their different cosmological dependencies, particularly their use as a
direct probe of inflation (e.g. Dodelson, Kinney \& Kolb, 1997;
Kamionkowski \& Kosowsky, 1998). With this in mind, a precise
measurement of the CMB polarization field would be
extremely useful and there are a number of upcoming experiments
designed to make such measurements of both the E-mode and, perhaps
more importantly, the odd-parity B-mode polarization signal. In addition
to planned satellite missions with the capability of large-scale,
high-precision polarization measurements (e.g. the Planck, Tauber
2004, and proposed Inflation 
Probe\footnote{http://universe.gsfc.nasa.gov/program/inflation.html} 
experiments), there are a number of funded ground-based experiments
designed to measure the polarization field with very high
signal-to-noise over a limited region of sky. In particular, the
QUaD\footnote{http://www.astro.cf.ac.uk/groups/instrumentation/projects/}
(QUEST\footnote{$Q$ and $U$ Extragalactic Submillimetre Telescope} and DASI), 
and BICEP\footnote{http://bicep.caltech.edu} (Background Imaging of 
Cosmic Extragalactic Polarization) experiments should begin
observations in 2005, while the CLOVER (CMB Polarization Observer;
Taylor et al. 2004) experiment is planned for 2008. 
A detection and measurement of B-mode polarization in the CMB would provide a
major step forward in the field and, if a B-mode signal were 
found to be present on large enough angular scales, this would constitute a
direct detection of primordial gravitational waves (Seljak \&
Zaldarriaga 1997; Kamionkowski, Kosowsky \& Stebbins 1997). Such a
measurement would have immense consequences for theories of the early
Universe, possibly providing a handle on the energy scale of inflation
(e.g. Kinney 1998). 

These ground-based experiments will measure the CMB signal over finite 
areas of sky, typically a few hundred square degrees. There already
exist well-established methods for recovering the true underlying CMB 
temperature power spectrum from observations of such finite areas of
sky and these are well tested (G\'orski 1994; Tegmark 1997; Bond, Jaffe \& Knox 1998; 
Dor\'e, Knox \& Peel 2001; Szapudi, Prunet \& Colombi 2001a; Szapudi et 
al. 2001b; Wandelt, Hivon \& G\'orksi 2001; Hansen, G\'orski \& Hivon
2002; Hivon et al. 2002). Although these methods, in principle, should
be equally applicable to the polarization power spectra, many have not
been tested with detailed simulations. Hansen \& G\'orski (2003)
investigate temperature and polarization power spectra estimation
using Gabor transforms while Chon et al. (2004) have extended the
analysis of Szapudi et al. (2001a) to construct a polarization power 
spectrum estimator based on correlation functions.

In this paper, we extend, and test with simulations of the CMB
temperature and polarization fields, the pseudo-$C_\ell$ method for
fast estimation of CMB power spectra. Although not optimal, the 
pseudo-$C_\ell$ method has the advantage that it is fast and
completely applicable to mega-pixel data sets such as those expected
from upcoming satellite and ground-based experiments. We have
generalised the method first introduced for the temperature field,
$T$, by Hivon et al. (2002), and extended to the $E$ and $B$ modes of 
polarization by Kogut et al. (2003), to the full set of six CMB power 
spectra: $C_{\ell}^{TT}$, $C_{\ell}^{EE}$, $C_{\ell}^{BB}$,
$C_{\ell}^{TE}$, $C_{\ell}^{TB}$ and $C_{\ell}^{EB}$. Note that this
method for recovering power spectra is completely applicable to other 
cosmological data sets such as large-scale weak lensing surveys, via
the transformation, $\{ T, Q, U, E, B\} \rightarrow 
\{ \mu, \gamma_1, \gamma_2, \kappa, \beta\}$, where $\mu$ is the
magnification, $\gamma_1$ and $\gamma_2$ are the two components of the
observed shear, $\kappa$ is the even-parity lensing convergence and
$\beta$ is a non-gravitational odd-parity distortion. 

We also present, and test with simulations, a new method for fast and 
accurate calculation, for any given cosmological model, of the full
covariance matrix of all six pseudo-$C_\ell$ power spectra. This
method will be extremely useful in estimating cosmological parameters
directly from the observed pseudo-$C_\ell$ power spectra. We also
present the analytic expressions for the covariances (see also the
calculations of Hansen \& G\'orski 2003 and Challinor \& Chon
2004). These expressions are not easily simplified and evaluated so we 
propose fitting the mixing matrices which appear in these expressions
from Monte-Carlo (MC) simulations. Provided that these mixing matrices
are mostly independent of cosmology, this method allows one to
calculate the covariance matrix with relative ease for any given
cosmological model.   

The paper is organised as follows. In Section 2, we present the
calculation of the pseudo-$C_\ell$ method relating the observed to
full-sky power and cross-power spectra for the temperature 
($T$) and polarization ($E,B$) fields of the CMB. In Section 3, we
calculate the full covariance matrix of the pseudo-$C_\ell$ estimator
 and we outline our method for calculating it for a given
cosmological model.  In Section 4, we test the pseudo-$C_\ell$
reconstructions and the covariance matrix fitting method on realistic
simulations of CMB temperature and polarization experiments for both
full-sky and finite-area data sets, and we also discuss the results
and implications of these simulation tests. Our conclusions are
presented in Section 5.
    
%%%%%%%%%%%%%%%%%%%%%%%%%%%%%%%%%%%%%%%%%%%%%%%%%%%%%%%%%%%%%%%%%%%%%%
\vspace{-5mm}
\section{Pseudo $C_{\ell}$ Method for CMB power spectra}
\label{pclmethod}

As is well known, in the currently favoured inflationary based 
cosmological model, the density perturbations are expected to be
isotropic and nearly Gaussian distributed (e.g. Liddle \& Lyth 2000). 
In this case, all the information present in the CMB temperature and 
polarization fields can be compressed without loss into the power
spectra of the fields. The extraction of these power spectra from
pixelized and cut-sky maps has become a real challenge when analysing 
large CMB data sets. Indeed, the standard approach, based on maximum 
likelihood methods, either in Fourier sapce (Gorski 1994; Gorski et
al. 1994) or real space (Bunn 1995; Bond et al. 1998), is
extremely CPU intensive for a large number of pixels. 
The need for faster methods has led to the
development of a number of alternative techniques for power spectrum
estimation. Szapudi et al. (2001) and Chon et al. (2004) have
developed power spectrum estimators based on correlation functions
while Dor\'e et al. (2001) have proposed a maximum likelihood
estimator based on hierarchical decomposition of a CMB map into sub-maps
with varying degrees of resolution. Hansen et al. (2002) and Hansen \&
G\'orski (2003) use Gabor transforms to recover power spectra from
apodized regions of the sky, while recently,  Wandelt, Larson \&
Lakshminarayanan (2004) have proposed an exact method for recovery of
power spectra from CMB data using Gibbs sampling. An overview of power
spectrum estimators for the CMB temperature field can be found in
Efstathiou (2004).

The application of the pseudo-$C_\ell$ method to CMB observations was
first proposed by Wandelt et al. (2001) and Hivon et al. (2002). This
technique, which is based on the original method of Peebles (1973), 
relates the observed power spectra, $\widetilde{C}_\ell$, to the true 
underlying spectra, $C_\ell$, via a coupling matrix which describes
the effect of the sky-cut applied to the data. In the following
section we begin by deriving analytical expressions for
the full set of two-point self-correlations and cross-correlations 
between the temperature, $T$, and polarization, $E$ and $B$, on a
finite area of the sky. As we expect theoretically the
cross-correlations between different parity fields to be zero ($B$ has
the opposite parity of $T$ and $E$), any non zero measurement of
$C_{\ell}^{TB}$ or $C_{\ell}^{EB}$ would be indicative of residual
noise left in the data, providing a powerful test of systematic
effects in the experimental pipeline. Alternatively, if one is
confident that systematic effects are small with respect to any
measurement of $C_{\ell}^{TB}$ or $C_{\ell}^{EB}$, then these spectra
can be used to test for parity violations (e.g. Lepora 1998; Lue, Wang 
\& Kamionkowski, 1999).  

%--------------------------------------------------------------------%

\subsection{The CMB temperature and polarization fields}
The CMB radiation is completely characterised by its temperature
anisotropy, $T$, and polarization, $P$, in each direction in the sky. 
The temperature may be conveniently expanded in spherical harmonics, 
\be
T(\Omega)=\sum_\lm T_\lm Y_\lm(\Omega)
\ee
with the inverse transform,
\be 
T_\lm = \int \! d \Omega\, T(\Omega) \, \Ylm^* (\Omega),
\label{eq:inv_t} 
\ee
where the integration is taken over the whole sky. As $T$ is a real
field, the harmonic modes obey the hermicity relation, $T^*_\lm =
(-1)^m T_{\ell,-m}$.

The polarization field can be described by the Stokes parameters, $Q$
and $U$, with respect to a particular choice of a coordinate system on the
sky (in relation to which the linear polarization is defined). One can 
conveniently combine the Stokes parameters into a single complex
quantity representing the polarization, $P=Q + iU$. Due to its
rotation properties, $P \rightarrow Pe^{i2\phi}$, 
the polarization is a spin-2 field (see Zaldarriaga \& Seljak 1997). 
One may thus expand $P$ and its complex conjugate in terms of the
spin-weighted spherical harmonics, $\sYlm$ (where $s$ denotes the spin), as
\ba 
P (\Omega)  &=& Q(\Omega) + i U(\Omega) \nn
	    &=& \sum_\lm (E_\lm + iB_\lm)\tYlm(\Omega) \nn 
P^*(\Omega) &=& Q(\Omega) - i U(\Omega) \nn 
            &=& \sum_\lm (E_\lm - iB_\lm)\mtYlm(\Omega),  
\ea 
where the summation in $m$ is over $-\ell \le m \le \ell$. The 
$E_\lm$ and $B_\lm$ are the spin-2 harmonic modes of the so-called
electric (i.e. `gradient') and magnetic (i.e. `curl') components of
the polarization field respectively. Using the orthogonality of the
spin-s spherical harmonics over the whole sphere,
\ba
\int \! d \Omega\, \sYlm(\Omega) \sYlmd^{*}(\Omega) = \delta_{\ell \ld} \delta_{m m'},
\ea 
where the spin states must be equal, the harmonic modes of the $E$ and
$B$ fields can be found directly from the polarisation field, $P$; 
\[
E_\lm = \frac{1}{2} \int \! d \Omega\, [P (\Omega)\tYlm^*(\Omega) + 
P^* (\Omega)\mtYlm^*(\Omega)]
\]
\be
B_\lm = \! \frac{-i}{2} \! \int \! d \Omega\,[P(\Omega)
\tYlm^*(\Omega) - P^*(\Omega) \mtYlm^*(\Omega)].
\label{eq:inv_eb} 
\ee

%--------------------------------------------------------------------%

\subsection{Finite Area Fields}
In this paper, we are interested in the effects of limited sky
coverage on the measurement of the auto- and cross-power spectra of
the CMB temperature and polarisation fields. In the case of the
temperature field, the effect of a finite window function is
multiplicative,
\be
\widetilde{T}(\Omega)=W_T(\Omega)T(\Omega),
\ee
where $W_T(\Omega)$ defines a weighting function over the whole sky,
between $0$ and $1$, and $\widetilde{T}$ is the temperature field 
on the cut sky. Outside of the survey, $W_T=0$. The actual value of the
window will depend on the survey scan strategy, foregrounds, etc. The
effect of the survey window function on the harmonic modes is a
convolution, 
\be
\widetilde{T}_\lm =\sum_{\lmd} \!_0W^\mmd_\lld T_\lmd,
\ee
where we have defined the harmonic space window function, 
\be
\!_0W^\mmd_\lld = \int \! d\Omega \, _0Y_\lmd(\Omega)W_T(\Omega)
\,_0Y_\lm^*(\Omega).  
\label{eq:w_1} 
\ee
In the case of the polarization field, the effect of the window
function is
\be
\widetilde{P}(\Omega)=W_P(\Omega)P(\Omega),
\ee
where $\widetilde{P}$ is the cut-sky field. Note that the temperature
and polarization window functions will not, in general, be equal;
$W_T(\Omega) \ne W_P(\Omega)$. Using equations
(\ref{eq:inv_eb}), we find the cut-sky $\widetilde{E}$ and
$\widetilde{B}$ fields are given by (Lewis, Challinor \& Turok 2002)
\ba 
\widetilde{E}_\lm &=& \sum_{\lmd} (E_\lmd W^+_\llmm + B_\lmd W^-_\llmm) \nn 
\widetilde{B}_\lm &=& \sum_{\lmd} ( B_\lmd W^+_\llmm - E_\lmd W^-_\llmm), 
\ea 
where we have defined
\ba W^+_\llmm &=& \frac{1}{2}(_2W^\mmd_\lld + _{-2}W^\mmd_\lld) \nn 
W^-_\llmm &=& \frac{i}{2}(_2W^\mmd_\lld - _{-2}W^\mmd_\lld), 
\label{eq:w_2} 
\ea 
with the spin-weighted harmonic space window functions (for $s=\pm2$) given by 
\be 
\!_sW^\mmd_\lld = \int \! d\Omega \, \sYlmd(\Omega)W_P(\Omega)
\sYlm^*(\Omega).  
\label{eq:w_3} 
\ee
Thus, the effect of observing the CMB radiation over a
limited area of the sky induces a contamination of the electric signal
by the magnetic component and vice-versa. This was to be expected as
the electric/magnetic (or `gradient/curl') decomposition of the
polarization vector on a finite patch is non-unique (see Lewis et
al. 2002).

We can simplify these expressions for the temperature and polarization
fields by defining the vector field
\be
{\mathbf \Theta}_\lm = ( \Theta^0_\lm, \Theta^1_\lm, \Theta^2_\lm ) = ( T_\lm, E_\lm,
B_\lm ).
\ee
The cut-sky vector, $\widetilde{\mathbf \Theta}_\lm$, is then related to
the underlying whole-sky field, ${\mathbf \Theta}_\lm$, by
\be 
{\mathbf {\widetilde \Theta}}_\lm = \sum_{\lmd} {\mathbf W}^\mmd_\lld {\mathbf
\Theta}_\lmd.
\label{eq:pseudo_fields} 
\ee 
where the window matrix is given by 
\be 
{\mathbf W}^\mmd_\lld = 
    \left( 
      \begin{array}{ccc} W^0_\llmm  & 0 & 0
                      \\ 0 & W^+_\llmm  & W^-_\llmm 
                      \\ 0 &-W^-_\llmm  & W^+_\llmm
      \end{array} 
    \right),\,\, 
\label{eq:window_matrix}
\ee 
with $W^0_\llmm = \!_0W^\mmd_\lld$ given by equation (\ref{eq:w_1}).

%--------------------------------------------------------------------%

\subsection{Statistical properties of the CMB on the cut sky}
\label{stats_cutsky}

One can use equations~(\ref{eq:pseudo_fields}) and (\ref{eq:window_matrix})
for the pseudo-temperature and polarization fields to relate the
values of the six pseudo-power spectra of the cut-sky fields to their 
full-sky counterparts. Taking the expectation values over the sphere
of the harmonic coefficients of the full-sky vector field, 
${\mathbf \Theta}_\lm$, we obtain the full-sky power spectra which we can 
write as a matrix relation,
\be
{\mathbf C}_{\ell} = \frac{1}{2\ell+1}\sum_{m}\lgl {\mathbf \Theta}_\lm
{\mathbf \Theta}_\lm^\dag \rgl,
\ee
where 
\be
\lgl {\mathbf \Theta}_\lm {\mathbf \Theta}_\lm^\dag \rgl \! =
 \!\! \left( \!\!\! \begin{array}{ccc} 
\lgl T_\lm T^*_\lm \rgl & \lgl T_\lm E^*_\lm \rgl & \lgl T_\lm B^*_\lm
\rgl \\ 
\lgl E_\lm T^*_\lm \rgl & \lgl E_\lm  E^*_\lm \rgl & \lgl E_\lm
B^*_\lm \rgl \\ 
\lgl B_\lm T^*_\lm \rgl & \lgl B_\lm E^*_\lm \rgl & \lgl B_\lm B^*_\lm
\rgl 
\end{array} \!\!\! \right). \,\, 
\ee
For fields defined only over finite areas of the sky, we can use 
equation~(\ref{eq:pseudo_fields}) to obtain an expression coupling the 
cut-sky power spectra to the full-sky spectra,
\ba
{\mathbf {\widetilde C}}_{\ell} &=& \frac{1}{2\ell+1}\sum_{m}\lgl {\mathbf
{\widetilde \Theta} }_\lm {\mathbf {\widetilde \Theta}}_\lm^\dag \rgl \nn
& = & \frac{1}{2\ell+1}\sum_{\ld} \sum_{m m'}
{\mathbf W}^\mmd_\lld {\mathbf C}_\ld ({\mathbf W}^\mmd_\lld)^\dag,
\ea 
which, for clarity, can be written as
\be 
{\mathbf {\widetilde C}}_{\ell} = \sum_{\ell'} {\mathbf M}_{\ell \ld} 
{\mathbf C}_{\ld}. 
\label{pcl1} 
\ee 
In Appendix~\ref{appendixA}, we outline the derivation of the full
coupling matrix, ${\mathbf M}$. Armed with these relations, the true
auto- and cross-power spectra can be estimated from the pseudo-power
spectra by inverting equation~(\ref{pcl1}),
\be   
{\mathbf C}_\ell = \sum_\ld {\mathbf M}_{\ell \ld}^{-1}
\widetilde{\mathbf C}_\ld.
\label{pcl2} 
\ee
In Section \ref{pclsims}, we shall apply this relation to a set of
numerical simulations of the CMB. But before we proceed, we now turn
our attention to calculating the covariances between the
pseudo-$C_\ell$ spectra.  

%%%%%%%%%%%%%%%%%%%%%%%%%%%%%%%%%%%%%%%%%%%%%%%%%%%%%%%%%%%%%%%%%%%%%%
\vspace{-5mm}
\section{Covariance matrix of the Pseudo-$C_\ell$ estimators}
\label{pclcovar}

In addition to estimating the full-sky temperature and polarization 
power spectra from the cut-sky spectra, we would also like to
understand the covariances between the different pseudo-$C_\ell$'s for 
model fitting or parameter estimation. In the case of maximum
likelihood parameter estimation, the covariance matrix is required at 
each point in parameter space sampled by the parameter estimation
procedure. Calculating it from MC simulations is then clearly
unfeasible if one wishes to investigate a large number of parameters. 
We therefore require an accurate and fast method for calculating the
covariance matrix of the pseudo-$C_\ell$ measurements. 

The covariance matrix of the various pseudo-$C_\ell$ spectra can be
expressed exactly in terms of the components of the window matrix,
equation~(\ref{eq:window_matrix}). Defining the covariance of two
measurable quantities, $X$ and $Y$, as 
\be
\lgl \Delta X \Delta Y \rgl = \lgl ( X - \lgl X \rgl )( Y - \lgl Y
\rgl ) \rgl,
\ee
where the angled brackets denote an ensemble average, the covariance
of the pseudo-$C_\ell$ spectra, for Gaussian-distributed temperature
and polarization fields, is given by (c.f. Hansen \& Gorski 2003;
Challinor \& Chon 2004)
\[
\lgl \Delta \widetilde{C}^{XY}_\ell \Delta \widetilde{C}^{MN}_\ld \rgl =
\]
\ba
\sum_{\ell_1 \ell_2} \left( C^{AD}_{\ell_1} C^{BC}_{\ell_2} X^{\{XA,
ND, MC, YB\}}_{\lld \ell_1 \ell_2} \right. + \nn
\left.  C^{AC}_{\ell_1} C^{BD}_{\ell_2} X^{\{XA, MC, ND, YB\}}_{\lld 
\ell_1 \ell_2} \right),
\label{eq: covar_all}
\ea 
where the $X_{\lld \ell_1 \ell_2}$ terms are given by
\[
X^{\{XA, ND, MC, YB\}}_{\lld \ell_1 \ell_2} =
\frac{1}{(2\ell+1)(2\ld+1)} \,\, \times
\]
\be
\sum W^{XA}_{\ell \ell_1 m m_1} (W^{ND}_{\ld \ell_1 m'm_1})^*
W^{MC}_{\ld \ell_2 m' m_2} (W^{YB}_{\ell \ell_2 m m_2})^* \!. 
\label{eq: xmatrices}
\ee
Here, the final sum is over all $m, m', m_1$ and $m_2$ and a sum over 
all possible values of $A$, $B$, $C$ and $D$ is implied.
The $W^{XY}_{\ell \ld m m'}$ matrices are just the entries of the
window matrix, equation~(\ref{eq:window_matrix}). Explicitly, these
are given by (suppressing the subscripts for clarity): $W^{TT}=W^{0}$, 
$W^{EE}=W^{+}$, $W^{BB}=W^{+}$, $W^{EB}=W^{-}$ and
$W^{BE}=-W^{-}$. For all other combinations of $X$ and $Y$,
$W^{XY}=0$. Similar expressions to equations (\ref{eq: covar_all}) and
(\ref{eq: xmatrices}) also arise in the calculation of the covariance
matrix of cross-power spectra obtained from different channels of
the same experiment for the temperature-only case (Tristram et al. 2005). 
For the temperature-only case, equation (\ref{eq: covar_all}) reduces
to 
\be
\lgl \Delta \widetilde{C}^{TT}_\ell \Delta \widetilde{C}^{TT}_\ld \rgl
= 2 \sum_{\ell_1 \ell_2} C^{TT}_{\ell_1} C^{TT}_{\ell_2} 
X^{\{TT, TT, TT, TT\}}_{\lld \ell_1 \ell_2}.
\label{eq: covar_TT}
\ee
In the case of a full-sky data set with a narrow galactic cut applied, 
Efstathiou (2004) has proposed approximating this exact expression by
replacing $C^{TT}_{\ell_1}$ and $C^{TT}_{\ell_2}$ with $C^{TT}_{\ell}$
and $C^{TT}_{\ld}$ and applying the completeness relation for spherical 
harmonics (Varshalovich et al. 1988). This yields an expression for
the covariance matrix of the pseudo-$C_\ell$ measurements in terms of
the temperature coupling matrix of equation~(\ref{eq:
M_TT}). Unfortunately an equivalent approach for the
full set of pseudo-$C_\ell$ estimators does not result in a simple
expression for the covariance matrix since the completeness relation 
for spin spherical harmonics is only valid for spherical harmonics of
the same spin. Recently, Challinor \& Chon (2004) have performed
additional approximations yielding analytic expressions for a
restricted part of the pseudo-$C_\ell$ covariance matrix in terms of
the Wigner-3j symbols. 
  
Here we propose a different method for calculating the covariances of
equation~(\ref{eq: covar_all}). Generalising the approach of
Efstathiou (2004), we make the symmetrized approximation,
\be
C^{AD}_{\ell_1} C^{BC}_{\ell_2} \rightarrow 
\sqrt{ C^{AD}_{\ell} C^{AD}_{\ld} C^{BC}_{\ell} C^{BC}_{\ld}},
\label{eq:symm_approx}
\ee
(and similarly for the second term in equation~\ref{eq: covar_all}) and
then fit the resulting $C_\ell$-independent terms to the covariance
measured from MC simulations. That is, after applying the above
approximation we are left with an equation of the form (where,
again, summation over values of $A$, $B$, $C$ and $D$ is implied),
\[
\lgl \Delta \tilde{C}^{XY}_\ell \Delta \tilde{C}^{MN}_\ld \rgl 
\approx  
\]
\ba
\sqrt{ C^{AD}_{\ell} C^{AD}_{\ld} C^{BC}_{\ell} C^{BC}_{\ld}}
\,\,\, {\bf X}_{\ell \ld}^{\{XA, ND, MC, YB\}} + \nn
\sqrt{ C^{AC}_{\ell} C^{AC}_{\ld} C^{BD}_{\ell} C^{BD}_{\ld}}
\,\,\, {\bf X}_{\ell \ld}^{\{XA, MC, ND, YB\}},
\label{eq: covar_fit}
\ea
where the $C_\ell$-independent mixing terms are given by
\be
{\bf X}_{\lld}^{\{XA, ND, MC, YB\}} = \sum_{\ell_1 \ell_2} 
X^{\{XA, ND, MC, YB\}}_{\lld \ell_1 \ell_2}.
\ee
For the covariances of the temperature pseudo-$C_\ell$'s only, we find
\be
\lgl \Delta \widetilde{C}^{TT}_\ell \Delta \widetilde{C}^{TT}_\ld \rgl
= 2 C^{TT}_{\ell} C^{TT}_{\ld} {\bf X}^{\{TT, TT, TT, TT\}}_{\lld},
\label{eq: covar_TT_new}
\ee
and in the case where the weighting function, $W_T(\Omega)$, is a simple 
mask with values of $1$ or $0$, the ${\bf X}_{\lld}$ matrix simplifies to
(Efstathiou 2004):
\be
{\bf X}^{\{TT, TT, TT, TT\}}_{\lld} = \frac{1}{2 \ld + 1}
M^{TT,TT}_{\lld},
\ee
where $M^{TT,TT}_{\lld}$ is the temperature coupling matrix of
equation~(\ref{eq: M_TT}). Unfortunately, the remaining 20 terms of
the ${\bf X}$-matrix do not simplify so easily. We provide the
explicit expressions for the mixing matrices and the 21 possible
covariances in Appendix~\ref{appendixB}. Note that the mixing matrices
exhibit some symmetries, in paticular, ${\bf X}^{abcd}_{\ell \ld} =
{\bf X}^{dcba}_{\ell \ld}$. While we could proceed to try
and find approximations to the ${\bf X}$-matrix elements, this would
introduce further uncertainties to the estimation of the covariances.
Instead, we propose fitting the ${\bf X}$-matrix terms to MC simulations of
the survey under consideration. Our assumption is that the mixing terms, 
${\bf X}_{\ell \ld}$, are very nearly independent of cosmology and
predominantly dependent on the sky-cut applied to the data. Note that
this same assumption is also required if we were to try and find
analytic approximations to the ${\bf X}_{\ell \ld}$ terms. We test 
this assumption and our proposed fitting method with simulations in
Section \ref{covar_test}. 

%%%%%%%%%%%%%%%%%%%%%%%%%%%%%%%%%%%%%%%%%%%%%%%%%%%%%%%%%%%%%%%%%%%%%%
\section{Tests on simulations}
\label{pclsims}

Equations~(\ref{pcl2}) and (\ref{eq: M_TT}) - (\ref{eq: M_EB})
provide us with a method for
recovering the various underlying full-sky CMB power spectra, ${\bf C}_{\ell}$,
from a set of pseudo-power spectra, $\widetilde{\bf C}_{\ell}$,
measured from a cut-sky map. In addition, the method outlined in
Section \ref{pclcovar} provides us with an approximation to the
pseudo-$C_\ell$ covariances for any given cosmological model. 
In this section, we test these procedures on simulated CMB temperature
and polarization maps. We first turn to the $\widetilde{\bf C}_\ell
\rightarrow {\bf C}_\ell$ reconstruction, after which we investigate
the covariance matrix approximation.  

%--------------------------------------------------------------------%

\subsection{Power spectra reconstructions}
\label{pcl_sims}

%--------------------------------------------------------------------%

\subsubsection{Simulated CMB maps}
\label{pclsims_sims}

We have used the 
{\sevensize HEALP}ix\footnote{available from http://www.eso.org/science/healpix} 
package (G\'orski, Hivon \& Wandelt 1999), \emph{synfast} to create
simulations of the temperature (T) and polarization (Q,U) fields of
the CMB sky. We have considered two experiments: (1) the Planck experiment
consisting of two sky surveys over 14 months and (2) a combination of
the QUaD polarization experiment (Church et al. 2003) with the
4-year temperature data from the WMAP satellite. The pixelization
scheme used in all the simulations is characterised by the {\sevensize
HEALP}ix parameter, $N_{\rm side}$, where we have used a value of 
$N_{\rm side}=2048$, giving an average pixel area of $2.95$
arcmin$^2$. Using such a small pixel size presents one with some
computing speed and data storage issues, especially when it comes to
running large numbers of MC simulations. However, this value of
$N_{\rm side}$ is necessary to ensure that both the Planck and  QUaD
beams, with full width at half maximums (FWHM) of $5$ and $4.2$ arcmin
respectively, are adequately sampled. If this was not the case, then
the resulting temperature and polarization maps would not be strictly
bandwidth-limited and the ensuing calculation of the pseudo-$C_\ell$'s
from the maps would potentially be biased at high $\ell$. Uncorrelated
Gaussian random noise has been added to the simulations in order to
mimic the instrumental noise effects in the detectors. The sensitivity 
levels used are those quoted on the
Planck\footnote{http://www.rssd.esa.int/index.php?project=PLANCK} and 
WMAP\footnote{http://map.gsfc.nasa.gov} websites and, for the QUaD
experiment, in Bowden et al. (2004). In each case, we have combined
sensitivity levels from the different channels in the three
experiments. The experimental parameters we have used to create the
simulations are given in Table \ref{tab:exp_params}. 

\begin{table*}
\center
\caption{Experimental parameters used to create the simulated Planck,
WMAP and QUaD datasets. Note that, for the Planck simulations, we have
considered only the 4 lowest frequency channels of the Planck High
Frequency Instrument (HFI) and, for the WMAP simulations we have
included the Q, V and W bands only. The other bands of these
experiments will mainly be used to constrain foregrounds. Noise levels
($\Delta T/T$) are given separately for temperature and polarisation
in terms of the sensitivity ($1\sigma$) to intensity (Stokes I) and 
polarised intensity (Stokes Q and U) fluctuations, relative to the average
temperature of the CMB (2.73 K). These sensitivity levels are given per
{\sevensize HEALP}ix pixel where the pixelisation scheme we have used
is characterised by the {\sevensize HEALP}ix parameter, $N_{\rm
side}$. For Planck, these noise levels are the performance level
goals for 2 sky surveys (14 months). The noise levels for WMAP are
those expected after 4 years of observations and for QUaD, the noise
levels are those expected after 2 years observing. Note that WMAP
will, in fact measure polarisation, and QUaD will measure
temperature. However, in this paper, we are interested in the
combination of the WMAP temperature and QUaD polarisation measurements
and so do not quote these sensitivity levels.} 
\label{tab:exp_params}
\begin{tabular}{@{}lcccccccccc}

\hline
& \multicolumn{4}{|c|}{Planck HFI} & \multicolumn{3}{|c|}{WMAP}
& \multicolumn{2}{|c|}{QUaD} \\
\hline

Sky fraction ($f_{\rm sky}$) & \multicolumn{4}{|c|}{0.826} & \multicolumn{3}{|c|}{0.826}
& \multicolumn{2}{|c|}{0.0078} \\

Frequency (GHz) & 100 & 143 & 217 & 353 & 41 & 61 & 94 & 100 & 150 \\
 
Beam FWHM (arcmin) & 9.5 & 7.1 & 5.0 & 5.0 & 30 & 21 & 13 & 6.3 & 4.2 \\ 

$\Delta T/T \times 10^{5}$ (temperature) & 1.07 & 0.91 & 1.40 & 4.27 & 2.85 &
3.47 & 3.28 & --- & --- \\ 

$\Delta T/T \times 10^{5}$ (polarisation) & 2.14 & 1.73 & 2.85 & 8.67 & --- & ---
& --- & 0.36 & 0.32 \\ 

{\sevensize HEALP}ix parameter, $N_{\rm side}$ &
\multicolumn{4}{c|}{2048} & \multicolumn{3}{c|}{512} & \multicolumn{2}{c|}{2048} \\

Average pixel area (sq. arcmin) &
\multicolumn{4}{c|}{2.95} & \multicolumn{3}{c|}{47.2} & \multicolumn{2}{c|}{2.95} \\

\hline
\end{tabular}
\end{table*}

The simulations are random Gaussian realisations of
theoretical {\sevensize CMBFAST}\footnote{available from
http://www.cmbfast.org} (Seljak \& Zaldarriaga, 1996)
temperature and polarization power spectra. Note that we have used
a slightly modified version of {\sevensize CMBFAST} in which the
amplitude of the initial curvature perturbations is characterised by the
parameter, $A$. This is the same parametrization used by the WMAP team
(see Verde et al. 2003 for details). The cosmological parameters used to generate
the {\sevensize CMBFAST} spectra were the best-fitting parameters of
the WMAP experiment for their power law $\Lambda$CDM model (Spergel et
al. 2003):
\[
\{ \Omega_{\rm m}, \Omega_\Lambda, \Omega_{\rm b}, h, n_{\rm s}, \tau,
A \} \]
\be
\hspace{1cm} = \{ 0.27, 0.73, 0.046, 0.72, 0.99, 0.17, 0.90 \}. 
\ee 
Here, $\Omega_m$ is the total matter density, $\Omega_\Lambda$ is the
dark energy density, $\Omega_{\rm b}$ is the energy density in baryons,
$h$ is the Hubble constant in units of $100$ kms$^{-1}$Mpc$^{-1}$,
$n_{\rm s}$ is the spectral index of the initial perturbations, $\tau$
is the optical depth to the last scattering surface and $A$ is the
parameter used by the WMAP team to characterise the amplitude of the
initial perturbations.

\begin{figure*}
\centering
\begin{picture}(200,300) 
\includegraphics{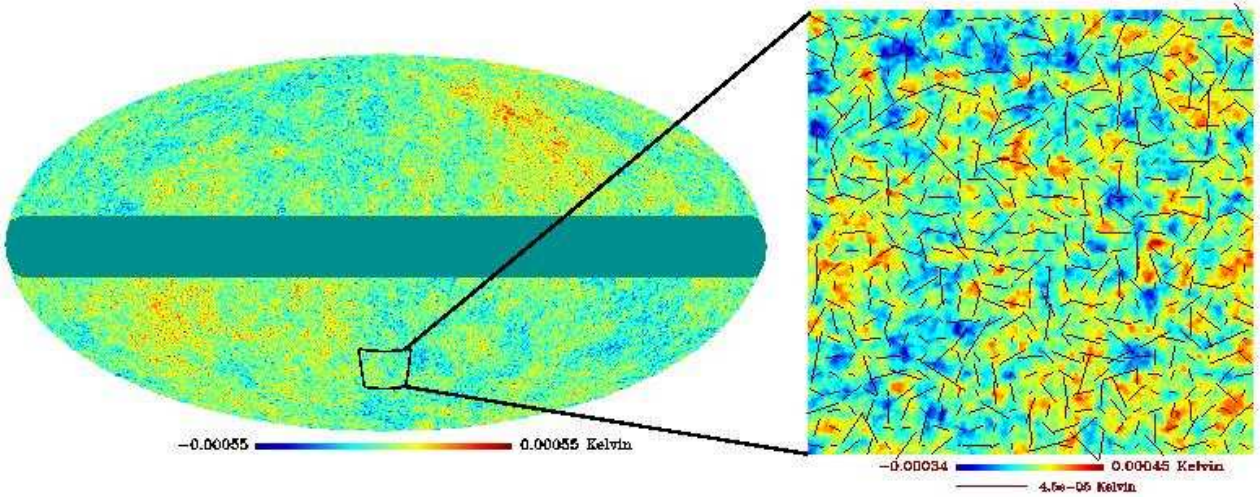}
\end{picture}
\vspace{-3.2cm}
\caption{Example simulated maps for the Planck (temperature - left)
and QUaD (polarization - right) experiments. A $\pm10^{\circ}$
galactic cut has been excised from the simulated Planck maps. The
temperature field is shown as the colour scale and the polarization
field in the QUaD simulation is represented by the vector field. The
orientation of each vector indicates the direction of polarization while
the length indicates the polarization magnitude. The simulated QUaD
map measures $17.5\times17.5$ square degrees - similar to the proposed
QUaD survey area (Bowden et al. 2004).}
\label{maps}
\end{figure*}

In addition, to test the reconstruction of the BB
power spectrum, we have included tensor perturbations in our
cosmological model. We have used a tensor-to-scalar ratio of 
\be
T/S=\frac{\Delta^2_T(k_0)}{\Delta^2_{\cal R}(k_0)} = 0.05,
\ee	
where $\Delta^2_{\cal R}(k_0)$ is the power spectrum of the initial curvature
perturbations and $\Delta^2_T(k_0)$ is the power spectrum of
gravitational waves from inflation -- both evaluated at $k_0=0.05 \, {\rm Mpc^{-1}}$.
We have also included the effects of gravitational lensing in our
cosmological model which increases the amplitude of the B-mode
polarization signal on small to medium scales. Multipoles from
$\ell=0$ to $\ell=3000$ were used in the map-making process. 
After generating the maps, they were smoothed with a Gaussian beam
with a FWHM of 13 (WMAP), 5 (Planck) and 4.2 (QUaD) arcmin. These
FWHM values correspond to the highest resolution channel for each
of the three experiments (see Table \ref{tab:exp_params}). Example maps of 
the temperature (Planck) and  polarization (QUaD) fields are shown in 
Fig.~\ref{maps}.  

%--------------------------------------------------------------------%

\subsubsection{Reconstructing the power spectra}
\label{pclsims_method}

To measure the pseudo-$C_\ell$ spectra, we use the {\sevensize
HEALP}ix package, \emph{anafast} to estimate 
\be
\widetilde{\bf C}_\ell = \frac{1}{2\ell+1}\sum_m 
\widetilde{\bf \Theta}_\lm \widetilde{\bf \Theta}^\dag_\lm,  
\ee
where $\widetilde{\bf \Theta}_\lm = (\widetilde{T}_\lm, \widetilde{E}_\lm,
\widetilde{B}_\lm)$ are the observed spherical harmonic
coefficients for the temperature and polarization fields. These 
have been calculated from the simulated maps using equations~(\ref{eq:inv_t})
and (\ref{eq:inv_eb}) with $T(\Omega)$ and  $P(\Omega)$ replaced with
${\widetilde T(\Omega)}$ and ${\widetilde P(\Omega)}$ respectively.
The integrals in these equations have been approximated as sums over
discrete pixels. For the tests we present here, we have used the
simplest possible weighting scheme: $W_T=1$ for observed pixels and
$W_T=0$ otherwise (and similarly for $W_P$). Note that the method of
Section \ref{pclmethod} also allows for more complicated window
functions, e.g. inverse noise weighting for each pixel. The
\emph{anafast} package exploits various symmetries of spherical
harmonics to reduce the computation of the $\widetilde{\bf \Theta}_\lm$
coefficients from a $N_{\rm pix} \ell^2_{\rm max}$ process to one
which scales as $N^{1/2}_{\rm pix} \ell^2_{\rm max}$ where $N_{\rm
pix}$ is the number of pixels in the map and $\ell_{\rm max}$ is the
maximum multipole to be measured. 

When it comes to applying the method of Section \ref{pclmethod} to
recover the CMB spectra from the pseudo-$C_\ell$'s measured from an
experiment with a large sky-cut, it turns out that
equations~(\ref{pcl2}) and (\ref{eq: M_TT}) - (\ref{eq: M_EB})
cannot be applied directly since the coupling matrix,
${\bf M}$ become singular and un-invertible for large sky-cuts.  
In this case, it is necessary to bin the power spectrum estimates and 
reconstruct the true CMB spectra in terms of bandpowers. Following
closely the temperature analysis of Hivon et al. (2002), 
we calculate full-sky CMB bandpowers (${\bf P}_b$) as
\be
{\bf P}_b=\sum_{b'}{\bf M}^{-1}_{bb'} \, \sum_\ell P_{b'\ell} \,
\widetilde{\bf C}_\ell,
\label{bandpowers}
\ee
where the subscript $b$ denotes the bands whose boundaries in
$\ell$-space are defined by $\ell^{(b)}_{\rm low} < \ell^{(b)}_{\rm
high} < \ell^{(b+1)}_{\rm low}$. $P_{b \ell}$ is a binning
operator in $\ell$-space defined by 
\be
P_{b \ell} = 
\left\{ \begin{array}{ll} 
\frac{1}{2 \pi} \frac{\ell(\ell+1)}{\ell^{(b+1)}_{\rm low} - 
\ell^{(b)}_{\rm low}}, & \mbox{  if  } 2 \le \ell^{(b)}_{\rm low} 
\le \ell < \ell^{(b+1)}_{\rm low} \\
0, & \mbox{  otherwise.  } \end{array} \right. 
\ee
In equation (\ref{bandpowers}), we have also introduced the binned
coupling matrix, ${\bf M}_{bb'}$ which is constructed from the
coupling matrix of Section \ref{pclmethod} using
\be
{\bf M}_{bb'} = \sum_\ell P_{b \ell} \sum_\ld {\bf M}_{\lld} Q_{\ld b'},
\ee
where $Q_{\ell b}$ is the reciprocal operator of $P_{b \ell}$,
\be
Q_{\ell b}=
\left\{ \begin{array}{ll}
\frac{2 \pi}{\ell(\ell+1)}, & \mbox{  if  } 2 \le \ell^{(b)}_{\rm low} 
\le \ell < \ell^{(b+1)}_{\rm low} \\
0, & \mbox{  otherwise.  } \end{array} \right. 
\ee
In this way, we approximate the true CMB spectra as being piece-wise
constant where the quantity  ${\bf P}_b \equiv \ell(\ell+1){\bf C}_\ell/2\pi$ is 
taken to be constant within each band. We can also include in the
coupling matrix a correction for the effects of the beam and finite 
pixel size,
\be
{\bf M}_{bb'} =  \sum_\ell P_{b \ell} \sum_\ld {\bf M}_{\lld} B^2_\ld Q_{\ld b'},
\ee
where $B_\ell$ is the combined window function describing the beam
smoothing and pixelization of the experiment. To calculate the 
Wigner 3j symbols, which are required for the coupling matrices of
Appendix \ref{appendixA}, we have used the publicly available 
{\sevensize SLATEC} fortran subroutine, {\sevensize DRC3JJ.f}. This routine uses
the algorithm of Schulten \& Gordon (1975) which makes use of both
forward and backward recurrence relations to maintain numerical 
stability. It is both rapid and accurate, even for large multipole
values. 
      
In order to calculate the errors and covariance of the resulting
bandpower measurements, we run a number ($\sim 100$) of MC
simulations. Note that we find very little difference in the
covariances measured from 50 and 100 MC simulations and therefore
conclude that our suite of 100 simulations is sufficient to yield a 
good estimate of the covariances. After measuring the ${\bf P}_b$
bandpowers from each of the simulations, their covariance is then
given by the run-to-run scatter among the simulations:
\be
\lgl \Delta {\bf P}_b \Delta {\bf P}_b'\rgl = 
\lgl({\bf P}_b - \overline{\bf P}_b)({\bf P}_b' - \overline{\bf P}_b')
\rgl_{MC},
\label{bandpowers_covar}
\ee
where $\overline{\bf P}_b$ denotes the average of each bandpower over
all simulation runs. Finally, the effects of the noise can be removed
from the final bandpower estimates by replacing equation~(\ref{bandpowers}) with
\be
{\bf P}_b=\sum_{b'}{\bf M}^{-1}_{bb'} \, \sum_\ell P_{b'\ell} \, 
(\widetilde{\bf C}_\ell-\lgl \widetilde{\bf N}_\ell \rgl_{MC}),
\label{bandpowers2}
\ee  
where $\lgl \widetilde{\bf N}_\ell \rgl_{MC}$ is the average
pseudo-power spectrum measured from a new set of 100 signal-free noisy 
MC simulations. An estimate of the full-sky noise power spectrum is 
then
\be
{\bf N}_b=\sum_{b'}{\bf M}^{-1}_{bb'} \, \sum_\ell P_{b'\ell} \, \lgl
\widetilde{\bf N}_\ell \rgl_{MC}.
\label{bandpowers_noise}
\ee

%--------------------------------------------------------------------%

\begin{figure*}
\centering
\begin{picture}(200,300)
\includegraphics{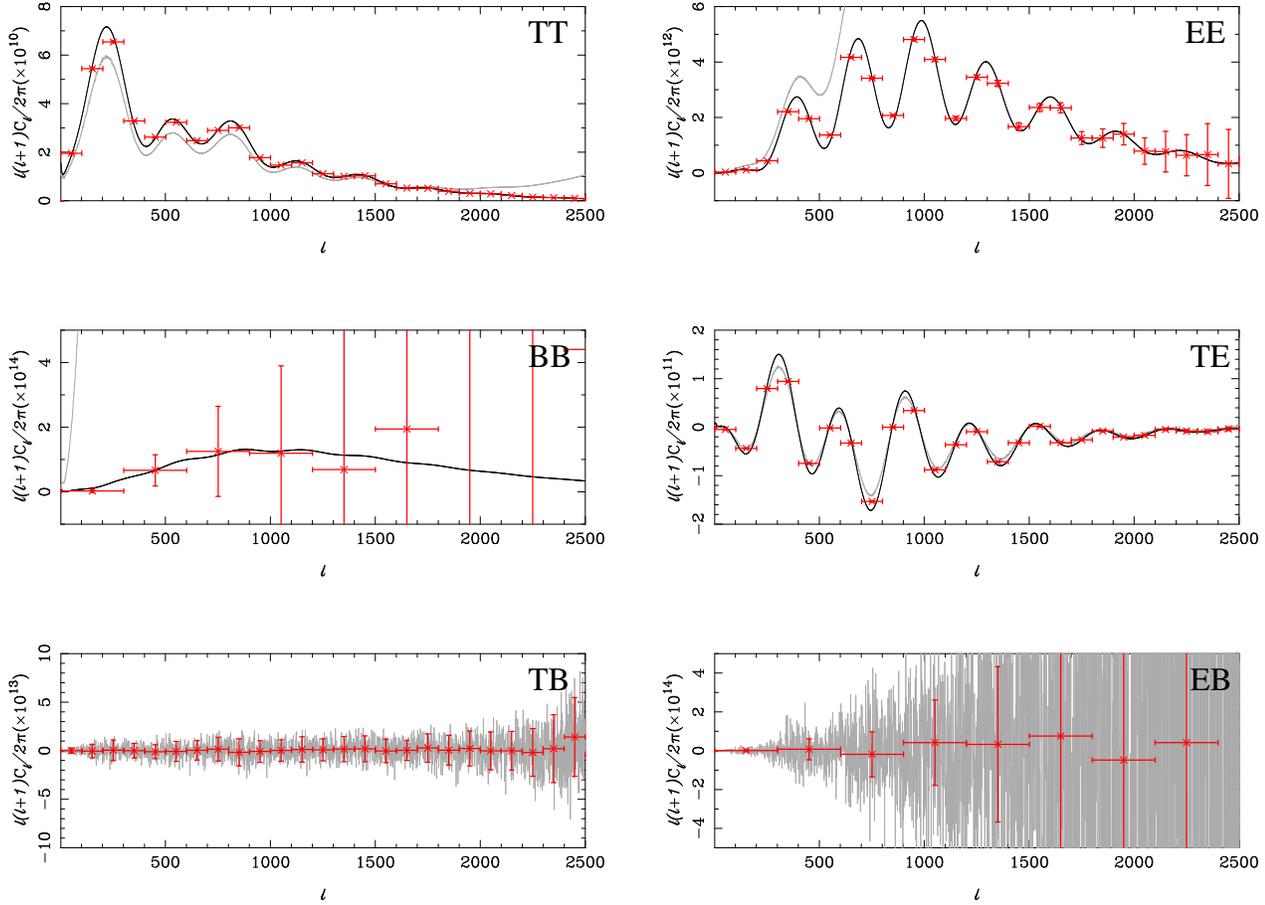}
\end{picture}
\vspace{2cm}
\caption{The CMB power spectra and cross power-spectra recovered in
terms of bandpowers from the simulated Planck data sets. The input
{\sevensize CMBFAST} spectra are shown as the black curves. The light 
grey curves are the pseudo-$C_\ell$ spectra as measured by the HEALpix 
package, \emph{anafast} and the points with error bars are the recovered
CMB bandpowers. The spectra are, clockwise from top-left: TT, EE, TE, 
EB, TB and BB.}
\label{planck_cl}
\end{figure*}

\subsubsection{Results}
\label{pclsims_results}

We have applied the $\widetilde{\bf C}_\ell \rightarrow {\bf C}_\ell$ 
reconstruction to the simulated Planck and QUaD + WMAP data sets in
terms of the bandpower formulation of the previous section. After
applying a $\pm10^{\circ}$ galactic cut to the simulated Planck and
WMAP maps, the pseudo-$C_\ell$ spectra were measured up to a maximum 
multipole of $\ell_{\rm max}=3000$. 
Equation~(\ref{bandpowers2}) was then applied to the measured 
$\widetilde{\bf C}_\ell$'s in order to recover the true underlying
full-sky power spectra. The reconstruction was carried out up to a
maximum multipole of $\ell_{\rm max}=3000$. However, for the very high 
$\ell$-modes, there will be significant contributions to the ${\bf
C}_\ell$ reconstruction from $\widetilde{\bf C}_\ell$ with $\ell >
3000$. We therefore only plot the recovered ${\bf C}_\ell$ up to a
maximum of $\ell=2500$, thus ensuring that contributions from
$\widetilde{\bf C}_\ell$ with $\ell > 3000$ are negligible. Note that
the resulting ${\bf C}_\ell$ measurements are also corrected for the
effects of the Gaussian beam and finite pixel size. 

In Fig.~\ref{planck_cl}, we plot the average recovered spectra
from our suite of Planck-like simulations. For these reconstructions,
we have used a linear binning scheme with bin sizes of $\Delta\ell=100$
for the TT, EE, TE and TB spectra and $\Delta\ell=300$ for the BB and EB
spectra. We use the coarser binning scheme for the latter two spectra
due to the lower significance of the signal for these spectra. The
error bars plotted are the $1\sigma$ errors calculated from the
diagonal elements of the covariance matrix
(equation~\ref{bandpowers_covar}). Note that we plot the errors
appropriate for a measurement from a single experiment. The error in
the mean recovered value, averaged over all the simulation runs, is
much smaller. We note that the recovery of the TT, EE and TE spectra
is unbiased over the entire range of multipoles plotted
($0\le\ell\le2500$) while the BB recovery is also unbiased with a
significant detection of BB power over a restricted range of
$\ell$-space. Comparing the reconstructed power spectra with the input
{\sevensize CMBFAST} models, we find reduced-$\chi^2$ values of 1.86, 0.83, 0.38
and 0.03 for the TT, TE, EE and BB spectra respectively. We suspect
that the relatively poor reduced-$\chi^2$ value of 1.86 for the
recovery of $C_\ell^{TT}$ from the Planck simulations is due to shot
noise from the finite number of simulations. This shot noise effect 
is only seen in the recovered Planck $C_\ell^{TT}$ because of the very low
temperature noise levels we have used in creating the Planck simulations.  
Finally we note that the non-cosmological TB and EB
spectra are consistent with zero over the entire multipole range.

\begin{figure*}
\centering
\begin{picture}(200,300) 
\includegraphics{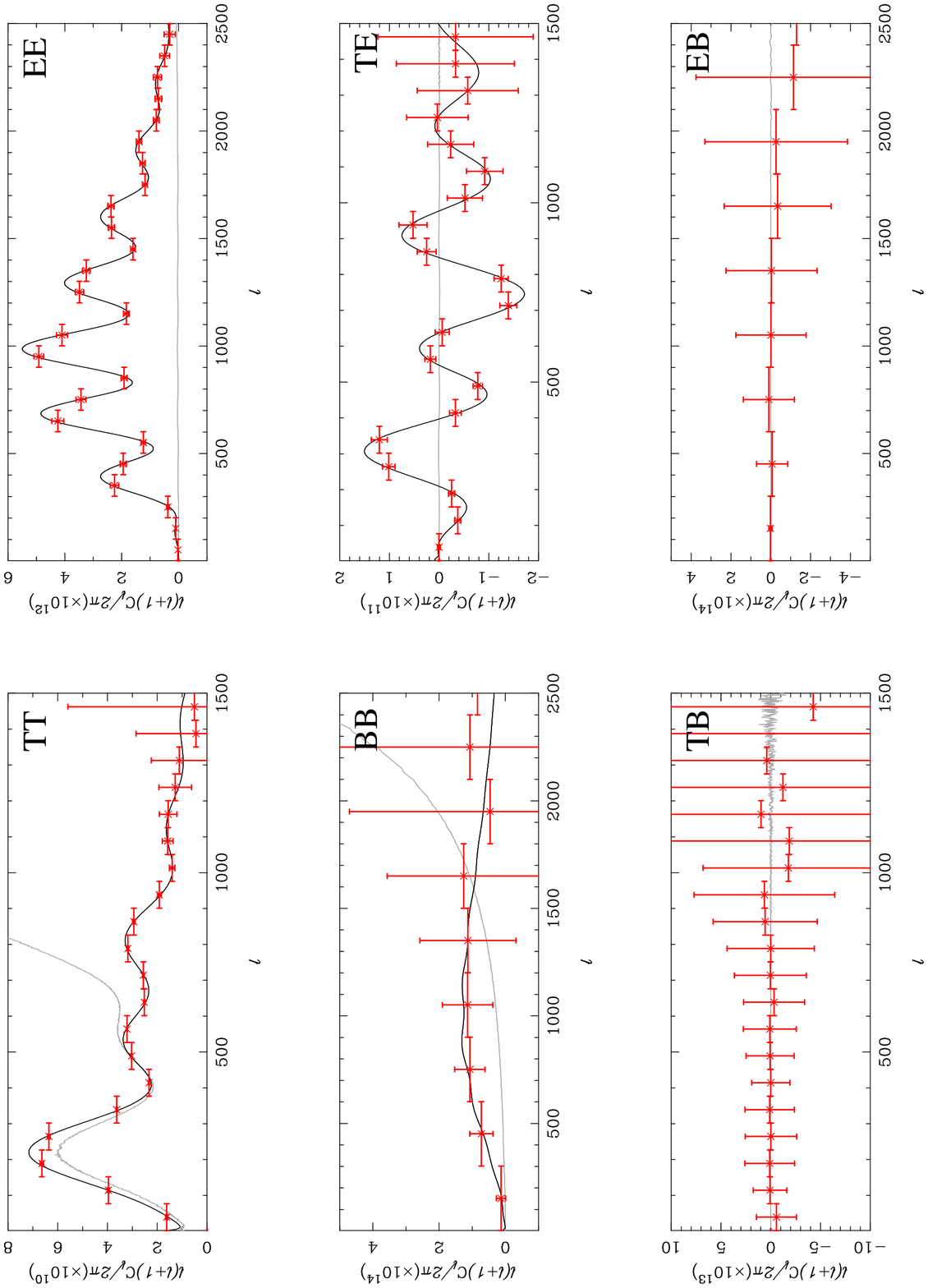}
\end{picture}
\vspace{2cm}
\caption{The CMB power spectra and cross power-spectra recovered in
terms of bandpowers from the simulated WMAP + QUaD data sets. 
The input {\sevensize CMBFAST} spectra are shown as the black curves. 
The light grey curves are the pseudo-$C_\ell$ spectra as measured by 
the HEALpix package, \emph{anafast} and the points with error bars are 
the recovered CMB bandpowers. The spectra are, clockwise from top-left: 
TT, EE, TE, EB, TB and BB.}
\label{qwmap_cl}
\end{figure*}
 
The recovered bandpowers for the combined QUaD + WMAP simulations are 
shown in Fig.~\ref{qwmap_cl}. Here, we plot the TT, TE and TB spectra up to
$\ell=1500$ only, as beyond this these spectra are unconstrained due to
the coarser resolution of the WMAP simulations (13 arcmin) compared to
that of Planck (5 arcmin). We have used a linear binning
of $\Delta \ell=75$ for the TT, TE and TB spectra, $\Delta \ell=100$
for the EE spectrum and $\Delta \ell=300$ for the BB and EB
spectra. Once again, the error bars plotted are the $1\sigma$
run-to-run errors calculated from the scatter among the simulations
(equation~\ref{bandpowers_covar}). We find excellent recovery of
the TT and TE spectra for $0\le\ell\le1300$. Beyond
$\ell=1300$, the effects of the 13 arcmin resolution of WMAP are
evident in these spectra. The recovery of the EE spectrum is excellent
over the entire range, $0\le\ell\le2500$, while for the BB spectrum,
the recovery is unbiased with a significant detection of the BB power
for $\ell\le1200$. These levels of detection for the polarization
spectra agree well with the predicted QUaD detection levels from the
Fisher matrix analysis of Bowden et al. (2004).  Again, the
non-cosmological TB and EB spectra are consistent with zero over the
entire multipole range. Comparing the recovered bandpowers with the input
spectra, we find reduced-$\chi^2$ values of 0.43, 0.03, 0.62 and 0.03
for the TT, TE, EE and BB spectra respectively. 

%--------------------------------------------------------------------%

\subsection{Testing the covariance matrix approximation}
\label{covar_test}
In Section \ref{pclcovar}, we have outlined our method for calculating
the full covariance matrix of the pseudo-$C_\ell$ estimators. Here, we
test our procedure against the covariance of the pseudo-$C_\ell$'s
measured from simulated data sets. Our procedure is to fit the 
${\bf X}_{\lld}$ matrices of equations~(\ref{eq:app2}) -
(\ref{eq:app22}) from the covariance measured from a number of MC
simulations. Our goal here is to reproduce the covariances of the
pseudo-$C_\ell$'s measured directly from the maps rather than the
covariances of the reconstructed bandpowers of the previous section. 
Note that, in some cases, simulations from at least three different 
cosmological models are required to find unique solutions for all the
mixing matrices. 

\subsubsection{Fitting procedure}
\label{covarsims_procedure}
To generate the fits, we run a number ($\sim 1000$) of 
simulations for each of 38 cosmological models whose input
cosmological parameters are drawn from within the WMAP $3\sigma$
region of parameter space surrounding the best-fit model for the WMAP
data (Spergel et al. 2003). For each of these cosmological models, we 
then measure the pseudo-$C_\ell$ covariances from the scatter among
the pseudo-$C_\ell$ spectra measured from the simulations and then use 
these covariances and equations~(\ref{eq:app2}) - (\ref{eq:app22}) to 
fit for the various ${\bf X}_{\lld}$ mixing matrices, knowing the
input model spectra used. We average these mixing matrices over the 38 
cosmological models to produce the final mixing matrices, which we
then use to calculate our semi-analytic covariances. 
In order to test the performance of the semi-analytic fitting method,
we conduct new sets of ($\sim 1000$) simulations using different input cosmological
parameters from those that were used to fit for the ${\bf X}_{\lld}$
matrices. We can then compare the covariance estimated from the
scatter of the pseudo-$C_\ell$'s measured from the simulations and the
covariance calculated using our fitting method. We perform these
tests for a number of cosmological models spanning the WMAP 1, 2 and
3$\sigma$ confidence regions of parameter space. We do this for both a
Planck-type experiment (with a $\pm10^{\circ}$ galactic cut excised 
from the simulated maps) and for the proposed survey design for QUaD
with a survey area of $\sim 300$ deg$^{2}$. Note that due to the large
number of simulations required, we have considered only multipoles
between $0\le\ell\le500$ as estimating the pseudo-$C_\ell$'s up to 
$\ell=2500$ from such a large number of maps is computationally
unfeasible at present. 

\begin{figure*}
\centering
\begin{picture}(200,300)
\includegraphics{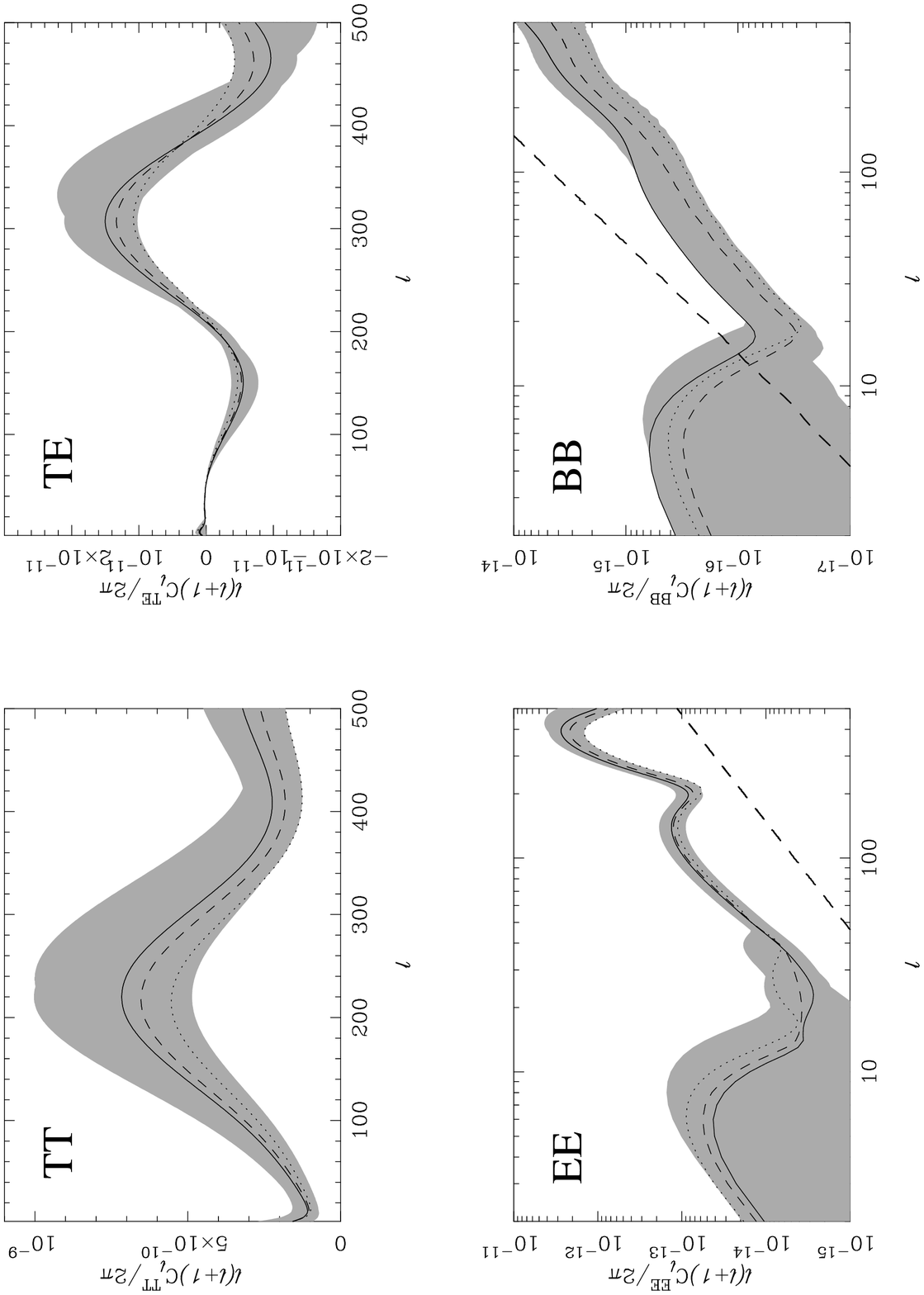}
\end{picture}
\vspace{2cm}
\caption{The range of CMB spectra used to perform the semi-analytic
fitting of the covariance matrix. The shaded regions show the $C_\ell$
range corresponding to the WMAP $3\sigma$ parameter space region 
used for the fits. The best-fit WMAP model is shown as the solid
curve. Also shown are two example comparison models taken from the WMAP $1$ 
(dashed) and $2\sigma$ (dotted) regions of parameter space which we use
to demonstrate the performance of the fitting method (see Section
\ref{covarsims_results}). The heavy dashed lines show the power
spectrum of the noise used in the simulations. Note that for the
Planck- and QUaD-like experiments which we have considered, the TT
noise power spectrum is negligible for $\ell<500$.}
\label{model_spectra}
\end{figure*}

The $C_\ell$-range corresponding to the ensemble of cosmological models we
have used to perform the fits are shown as the shaded regions in
Fig.~\ref{model_spectra}. Also indicated on these plots are the WMAP
default cosmological model and two comparison models drawn from the
$1$ and $2\sigma$ WMAP parameter space regions which we use to
assess the performance of our method (see Section
\ref{covarsims_results} below). 

%During the analysis, we noted that there were near
%degeneracies in many of the simultaneous equations we needed to solve 
%in order to fit for the ${\bf X_\lld}$ matrices. This resulted in some linear
%combinations of these equations being ill-conditioned which, in turn
%led to the fitted ${\bf X_\lld}$ matrices becoming unstable. To avoid
%this problem, we have used the singular valued decomposition (SVD,
%e.g. Press et al. 1992) technique to do the fitting which allows one
%to explicitly remove linear combinations of simultaneous equations
%which are ill-conditioned. The resulting solution obtained using this
%technique is the best-fit in the sense that it is the solution which 
%minimises the residuals, $|{\bf A}  {\bf x} - {\bf b}|$ when
%solving the matrix equation, ${\bf A}  {\bf x} = {\bf b}$. Using
%this method, the instabilities in the fitted ${\bf X_\lld}$ matrices
%were removed. 

A potential issue arises in the fitting procedure due to the fact that the
$C^{TE}_\ell$ power spectrum crosses zero and is negative in some
regions of $\ell$-space. A consequence of the symmetrized approximation, 
equation~(\ref{eq:symm_approx}), is the appearance of pre-factors
containing the quantity $\sqrt{C^{TE}_\ell C^{TE}_\ld}$ in some of the
covariance expressions (equations \ref{eq:app2} - \ref{eq:app22}). 
For some combinations of $\ell$ and $\ld$ then, we are
left with imaginary quantities for these pre-factors. 
To avoid this problem, we have simply replaced $\sqrt{C^{TE}_\ell
C^{TE}_\ld}$ by $\sqrt{|C^{TE}_\ell C^{TE}_\ld |}$ in equations
(\ref{eq:app2}) - (\ref{eq:app22}). In addition, for the covariances which 
depend on a single term containing $C^{TE}_\ell$, some of the mixing
matrices have discontinuities where $C^{TE}_\ell$ is very close to
zero, with resulting discontinuities in the fitted covariances. For the
small fraction of $(\ell,\ld)$-space where this occurs, we have
calculated the covariance by linearly interpolating from neighbouring 
parts of the fitted covariance matrix. 

\vspace{-0.4cm}
\subsubsection{Including noise}
\label{covarsims_noise}
For the tests we have conducted in this section, we have included
Gaussian random noise in the simulations with sensitivity levels
corresponding to those appropriate for the Planck and QUaD
experiments. To a good approximation, the effects of noise can be
accounted for in our fitting procedure by replacing the model power
spectra, $C_\ell^{XY}$ of equations~(\ref{eq:app2}) - (\ref{eq:app22})
with the sum of the cosmological plus noise power spectra,
$C_\ell^{XY}+N_\ell^{XY}$. Here, $N_\ell^{XY}$ are the full-sky
power spectra of the noise which we have evaluated from noise-only 
full-sky simulations. Alternatively one could use
equation~(\ref{bandpowers_noise}) to reconstruct the full-sky noise
power spectra from an experiment with limited sky coverage. 
Note that for the uncorrelated noise we have used in these tests,
$N_\ell^{TE}=N_\ell^{TB}=N_\ell^{EB}=0$. The power spectra of the
noise levels simulated are shown as the heavy dashed lines in 
Fig.~\ref{model_spectra}. Note that for both a Planck- and QUaD-like
experiment, $N_\ell^{TT}$ is negligible for $\ell < 500$.  

\subsubsection{Results}
\label{covarsims_results}
To display the results of the covariance comparisons, we use the
correlation matrix of the pseudo-$C_\ell$ measurements. Inserting the 
pseudo-$C_\ell$ spectra into the $6 \, \ell_{\rm max}$-element vector, 
$\widetilde{\bf C}_\ell$, which includes the six pseudo-spectra, in
order, TT, EE, BB, TE, TB and EB, then the $6 \, \ell_{\rm max} \times 
6 \, \ell_{\rm max}$ correlation matrix is defined by, 
\be
{\rm corr}(\widetilde{\bf C}_\ell,\widetilde{\bf C}_\ld) = 
\frac{ \lgl \Delta \widetilde{\bf C}_\ell \Delta \widetilde{\bf C}_\ld \rgl} 
{ \sqrt{ \lgl \Delta \widetilde{\bf C}_\ell \Delta \widetilde{\bf C}_\ell \rgl 
\lgl \Delta \widetilde{\bf C}_\ld \Delta \widetilde{\bf C}_\ld \rgl}},
\label{eq:corr_matrix}
\ee
which can take values between $-1$ (fully anti-correlated) and $+1$
(fully correlated) where the diagonal elements,
${\rm corr}(\widetilde{\bf C}_\ell,\widetilde{\bf C}_\ell)=1$ by design. 
Note that, for clarity, we plot the correlation matrix for multipoles 
in the range $1 \le \ell \le 100$ only. However, the levels of
agreement in this range are typical of the agreements found up to 
$\ell_{\rm max} = 500$. To investigate further the agreement between 
the true and semi-analytic covariances, we also plot the square root
of the diagonal elements of the covariance matrix (i.e. the 1$\sigma$
uncertainties) for each of the six pseudo-$C_\ell$ spectra.  We plot these
uncertainties up to the maximum multipole, $\ell_{\rm max}=500$ in each
case. 

\begin{figure*}
\centering
\begin{picture}(200,300)
\includegraphics{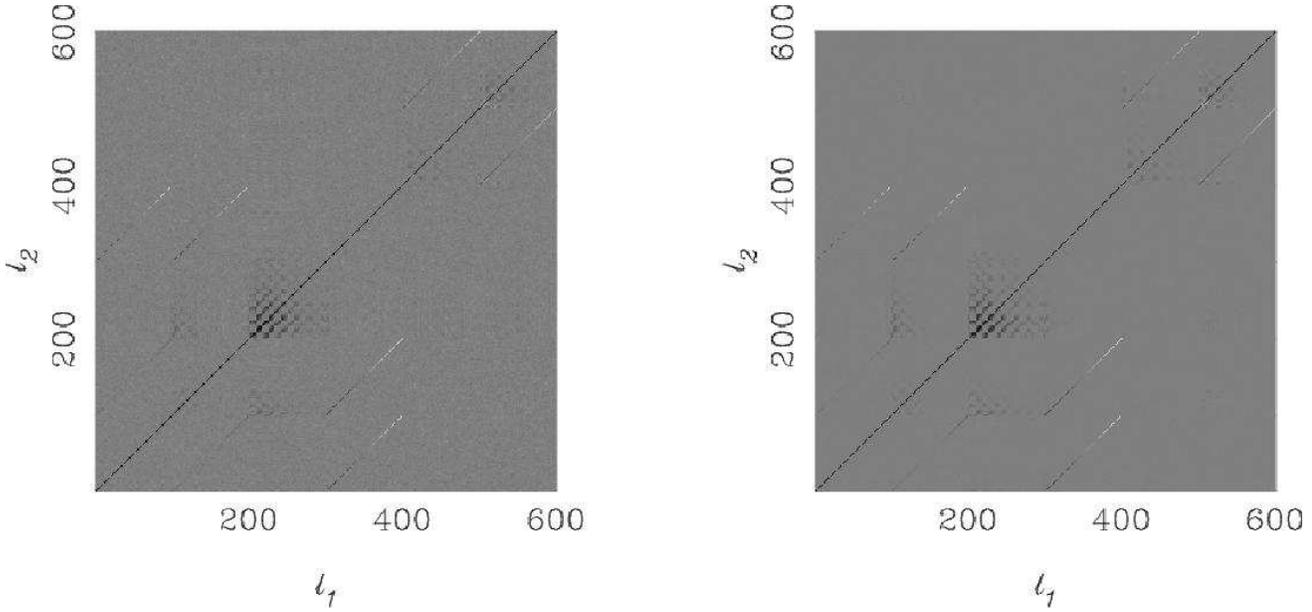}
\end{picture}
\vspace{-2.5cm}
\caption{The correlation matrix of the pseudo-$C_\ell$ measurements
from the Planck-like simulations (lhs) and using the semi-analytic 
approximation (rhs). The input cosmological model is taken from the
WMAP $1\sigma$ parameter space region. For clarity, we plot the
correlations for $1 \le \ell \le 100$ only. Multipoles in the range, 
$1\le\ell\le100$ are the TT spectra, $101\le\ell\le200$ are EE,
$201\le\ell\le300$ are BB, $301\le\ell\le400$ are TE,
$401\le\ell\le500$ are TB and $501\le\ell\le600$ are the EB
spectra.}
\label{cov_planck_1sigma}
\end{figure*}

\begin{figure*}
\centering
\begin{picture}(200,310)
\includegraphics{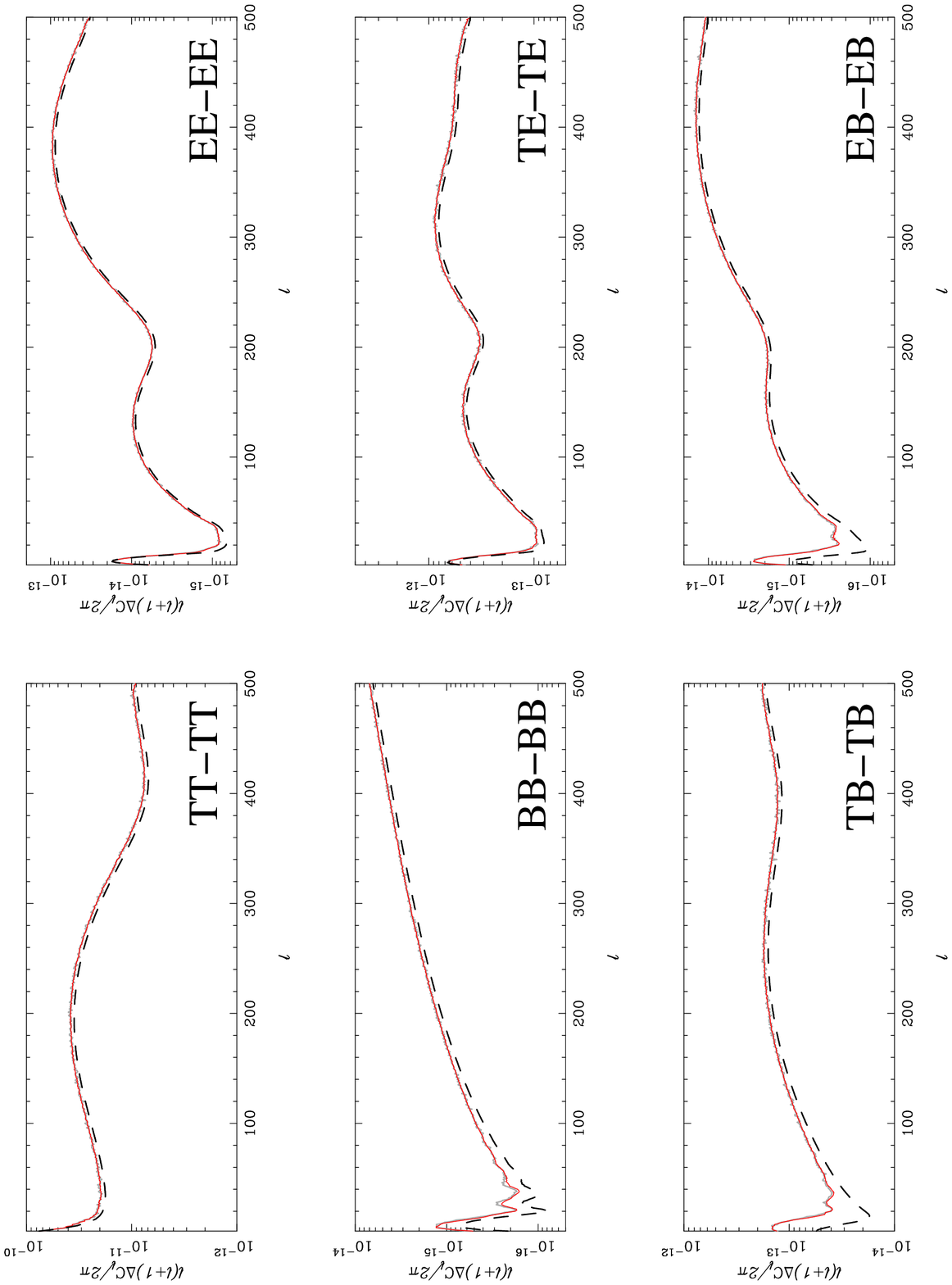}
\end{picture}
\vspace{1.6cm}
\caption{The square root of the diagonal elements of the covariance matrices depicted in
Fig.~\ref{cov_planck_1sigma}. In each plot, the light gray line is the
error measured from the simulations, the red (dark gray) line is the
semi-analytic approximation and the heavy dashed line is
the error calculated using the ``Knox'' formulae. The errors plotted are (clockwise
from top left) TT-TT, EE-EE, TE-TE, EB-EB, TB-TB and BB-BB.}
\label{diag_planck_1sigma}
\end{figure*}

Fig.~\ref{cov_planck_1sigma} compares the correlation matrix from
the Planck-like simulations and from the semi-analytic approximation
for a sample cosmological model from within the WMAP 1$\sigma$
parameter space region. The agreement between the correlation matrices is 
excellent with the semi-analytic approximation reproducing all the significant 
off-diagonal features seen in the covariance measured from the
simulations. Note that the covariance from the simulations also
contains shot-noise due to the finite number of simulations used which
is partly why the semi-analytic approximation appears smoother in these
plots. This effect, of course, is not important and would disappear
with an increase in the number of MC simulations used. The 1$\sigma$ errors
on the pseudo-$C_\ell$ estimates are plotted in
Fig.~\ref{diag_planck_1sigma}. Again, we find excellent agreement
between the simulations and the semi-analytic fits over the entire
multipole range. Once again, there are shot-noise effects in the
simulation curves. We have also tested the semi-analytic approximation
with simulations for cosmological models on the edge of the WMAP 2 and 
3$\sigma$ parameter space regions. For each of these regions, we have 
compared with 12 different cosmological models. As expected, the
further in parameter space one strays from the models used to generate
the fits, the worse the semi-analytic approximation becomes. Even so,
we found that this is quite a gradual effect and the approximation
is still performing well in the 3$\sigma$ region. This is encouraging
as it implies that the mixing matrices of equations~(\ref{eq:app2}) - 
(\ref{eq:app22}) are fairly insensitive to changes in the cosmological
model, at least for a modest sky-cut such as the $\pm 10^\circ$
galactic cut we have used in these tests.  

\begin{figure*}
\centering
\begin{picture}(200,300)
\includegraphics{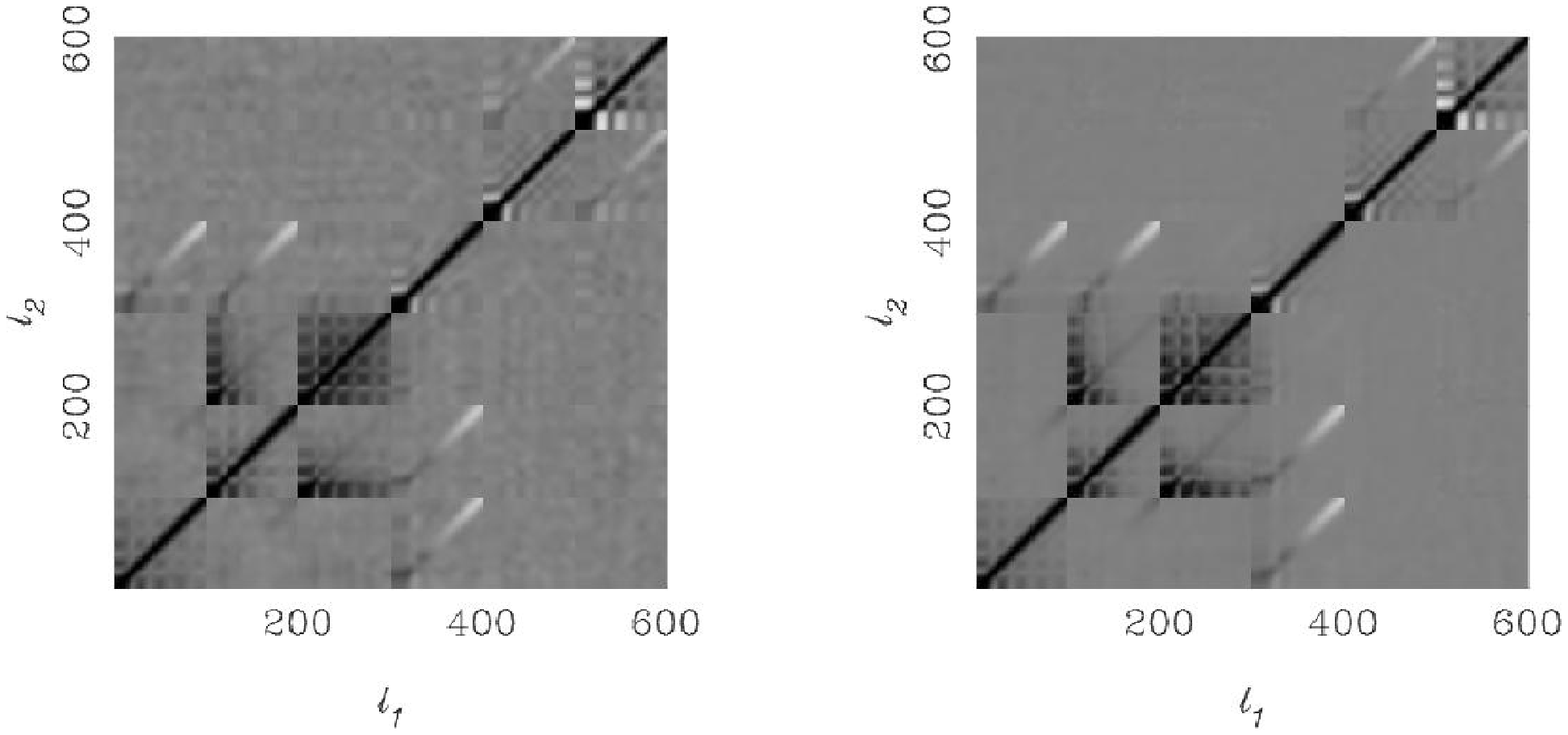}
\end{picture}
\vspace{-2.5cm}
\caption{The correlation matrix of the pseudo-$C_\ell$ measurements
from the QUaD-like simulations (lhs) and using the semi-analytic 
approximation (rhs). The input cosmological model is taken from the
WMAP $1\sigma$ parameter space region. For clarity, we plot the
correlations for $1 \le \ell \le 100$ only. Multipoles in the range, 
$1\le\ell\le100$ are the TT spectra, $101\le\ell\le200$ are EE ,
$201\le\ell\le300$ are BB, $301\le\ell\le400$ are TE,
$401\le\ell\le500$ are TB and $501\le\ell\le600$ are the EB spectra.}
\label{cov_quad_1sigma}
\end{figure*}

\begin{figure*}
\centering
\begin{picture}(200,310)
\includegraphics{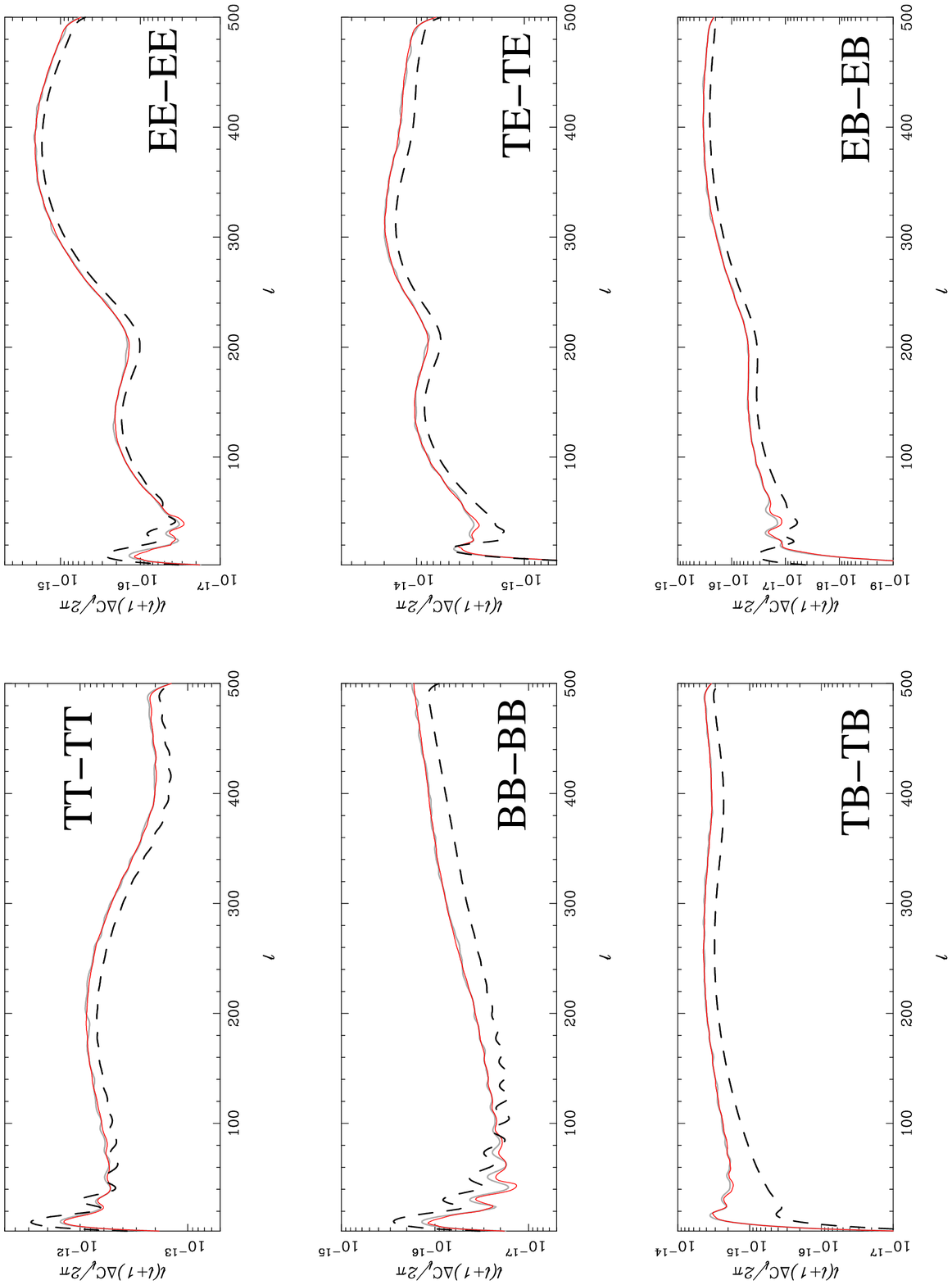}
\end{picture}
\vspace{1.6cm}
\caption{The square root of the diagonal elements of the covariance matrices depicted in
Fig.~\ref{cov_quad_1sigma}. In each plot, the light gray line is the
error measured from the simulations, the red (dark gray) line is the
semi-analytic approximation and the heavy dashed line is the
error calculated using the ``Knox'' formulae. The errors plotted
are (clockwise from top left) TT-TT, EE-EE, TE-TE, EB-EB, TB-TB and BB-BB.}
\label{diag_quad_1sigma}
\end{figure*}

\begin{figure*}
\centering
\begin{picture}(200,300)
\includegraphics{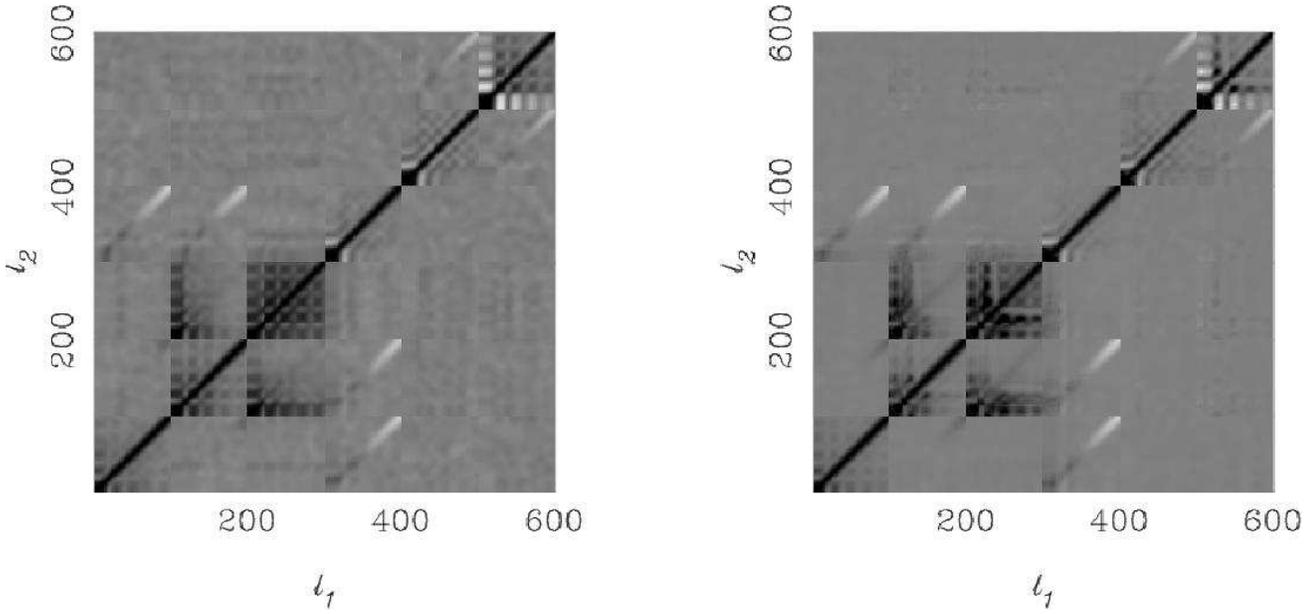}
\end{picture}
\vspace{-2.5cm}
\caption{The correlation matrix of the pseudo-$C_\ell$ measurements
from the QUaD-like simulations (lhs) and using the semi-analytic 
approximation (rhs). The input cosmological model is taken from the
edge of the WMAP $2\sigma$ parameter space region. Multipoles in the range, 
$1\le\ell\le100$ are the TT spectra, $101\le\ell\le200$ are EE ,
$201\le\ell\le300$ are BB, $301\le\ell\le400$ are TE,
$401\le\ell\le500$ are TB and $501\le\ell\le600$ are the EB spectra.}
\label{cov_quad_2sigma}
\end{figure*}

\begin{figure*}
\centering
\begin{picture}(200,310)
\includegraphics{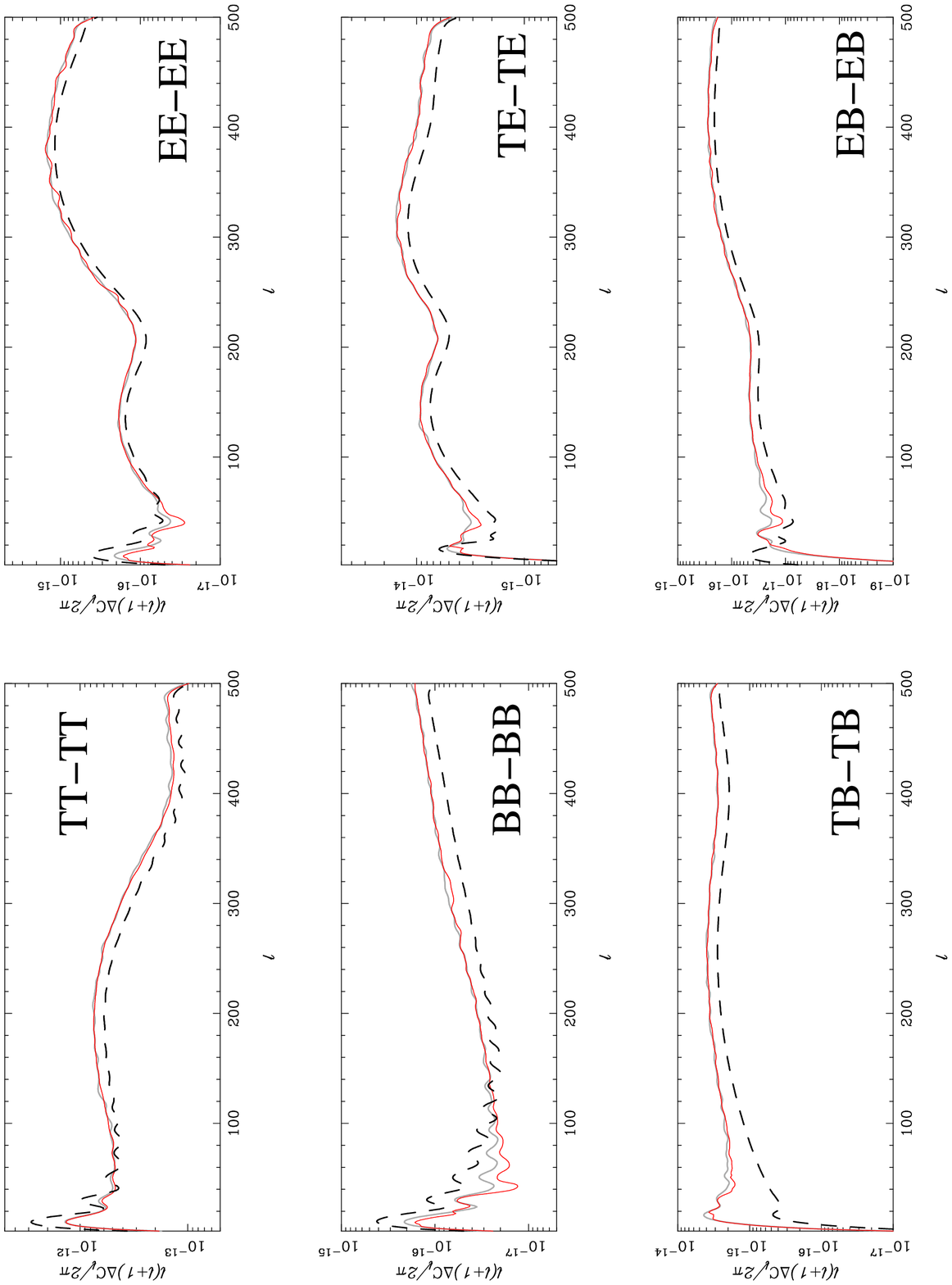}
\end{picture}
\vspace{1.7cm}
\caption{The square root of the diagonal elements of the covariance matrices depicted in
Fig.~\ref{cov_quad_2sigma}. In each plot, the light gray line is the
error measured from the simulations, the red (dark gray) line is the
semi-analytic approximation and the heavy dashed line is
the error calculated using the ``Knox'' formulae. The errors plotted are (clockwise
from top left) TT-TT, EE-EE, TE-TE, EB-EB, TB-TB and BB-BB.}
\label{diag_quad_2sigma}
\end{figure*}

The effect of a more drastic sky-cut can be seen in 
Figs.~\ref{cov_quad_1sigma} and \ref{diag_quad_1sigma} which show the
comparison between simulations and the semi-analytic approximation
(again for a sample model from the WMAP 1$\sigma$ parameter region) 
for the QUaD-like simulations with a survey area of $\sim 300$
deg$^2$. As expected, this sky-cut induces much stronger off-diagonal 
structure in the correlation matrices. Most of
this structure is reproduced in the semi-analytic approximation,
albeit slightly less accurately than for the case of the near full-sky
simulations. As in the case of the Planck-like simulations, we have
also compared with simulations from the WMAP 2 and 3$\sigma$
regions. Once again, we found the agreement gradually worsening as one 
moved away from the WMAP 1$\sigma$ region.  An example of the
performance of the approximation outside the WMAP $1\sigma$ region 
is given in Figs.~\ref{cov_quad_2sigma} and \ref{diag_quad_2sigma}
which show the correlation matrices and 1$\sigma$ errors respectively, for one
of the WMAP 2$\sigma$ region models, for the QUaD-like
simulations. The levels of agreement seen in
Figs.~\ref{cov_quad_2sigma} and \ref{diag_quad_2sigma} are typical of
the performance of the approximation outside of the WMAP $1\sigma$
region for a QUaD-like survey area, while for the Planck-like simulations,
the agreement is even better.  

\begin{figure*}
\centering
\begin{picture}(200,300)
\includegraphics{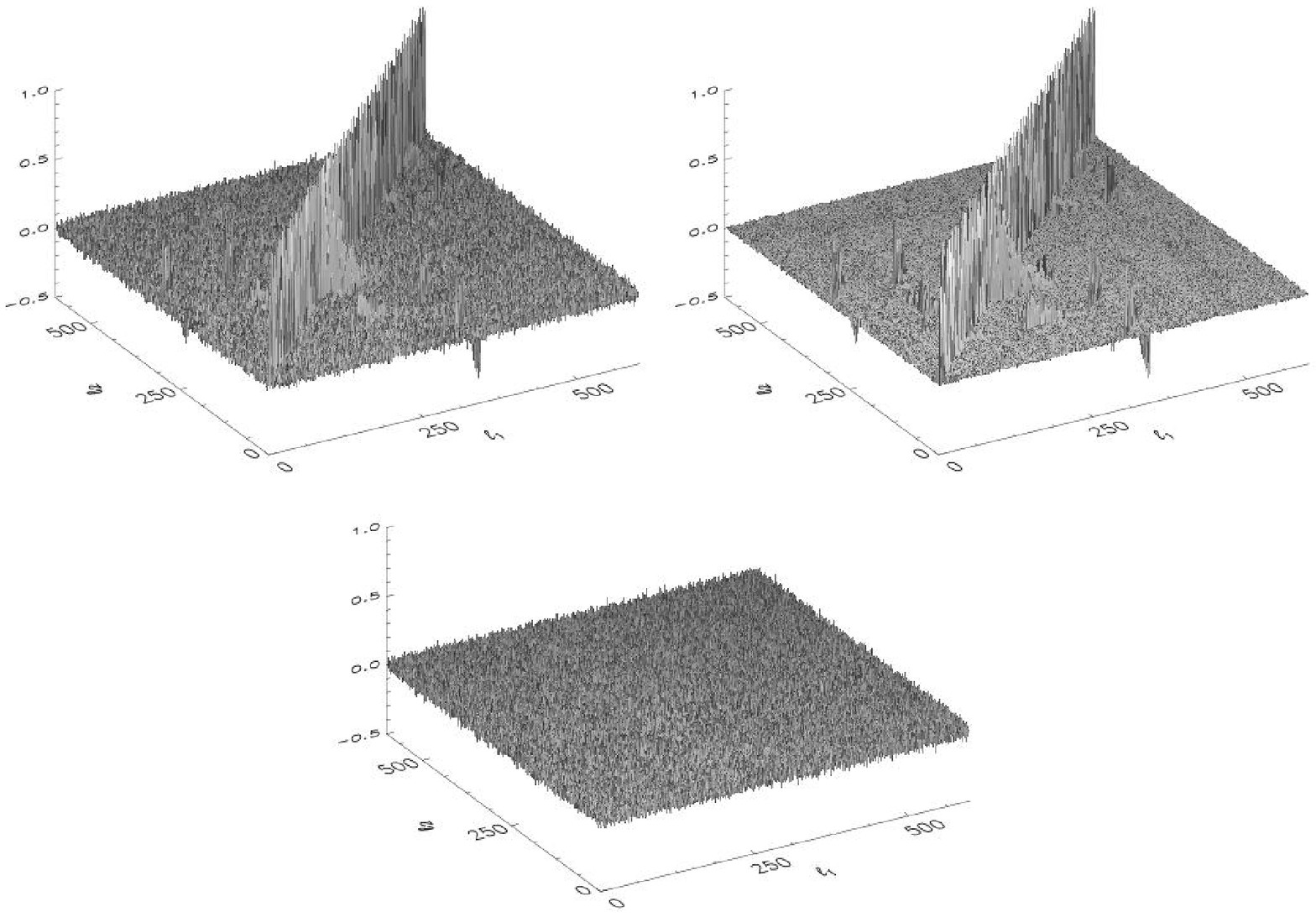}
\end{picture}
\vspace{1.5cm}
\caption{The simulated ({\it upper, left}) and semi-analytic ({\it upper,
right}) correlations for the WMAP 1$\sigma$ region Planck simulations
shown in Figs.~\ref{cov_planck_1sigma} and \ref{diag_planck_1sigma}. The
residuals between the two, which have an \emph{rms} value of $\sim 0.03$, are
shown in the lower plot. Once again, for clarity, we plot the
correlations for $1 \le \ell \le 100$ only. Multipoles in the range, 
$1\le\ell\le100$ are the TT spectra, 
$101\le\ell\le200$ are EE , $201\le\ell\le300$ are BB,
$301\le\ell\le400$ are TE, $401\le\ell\le500$ are TB and
$501\le\ell\le600$ are the EB spectra.}
\label{residuals_planck}
\end{figure*}

\begin{figure*}
\centering
\begin{picture}(200,300)
\includegraphics{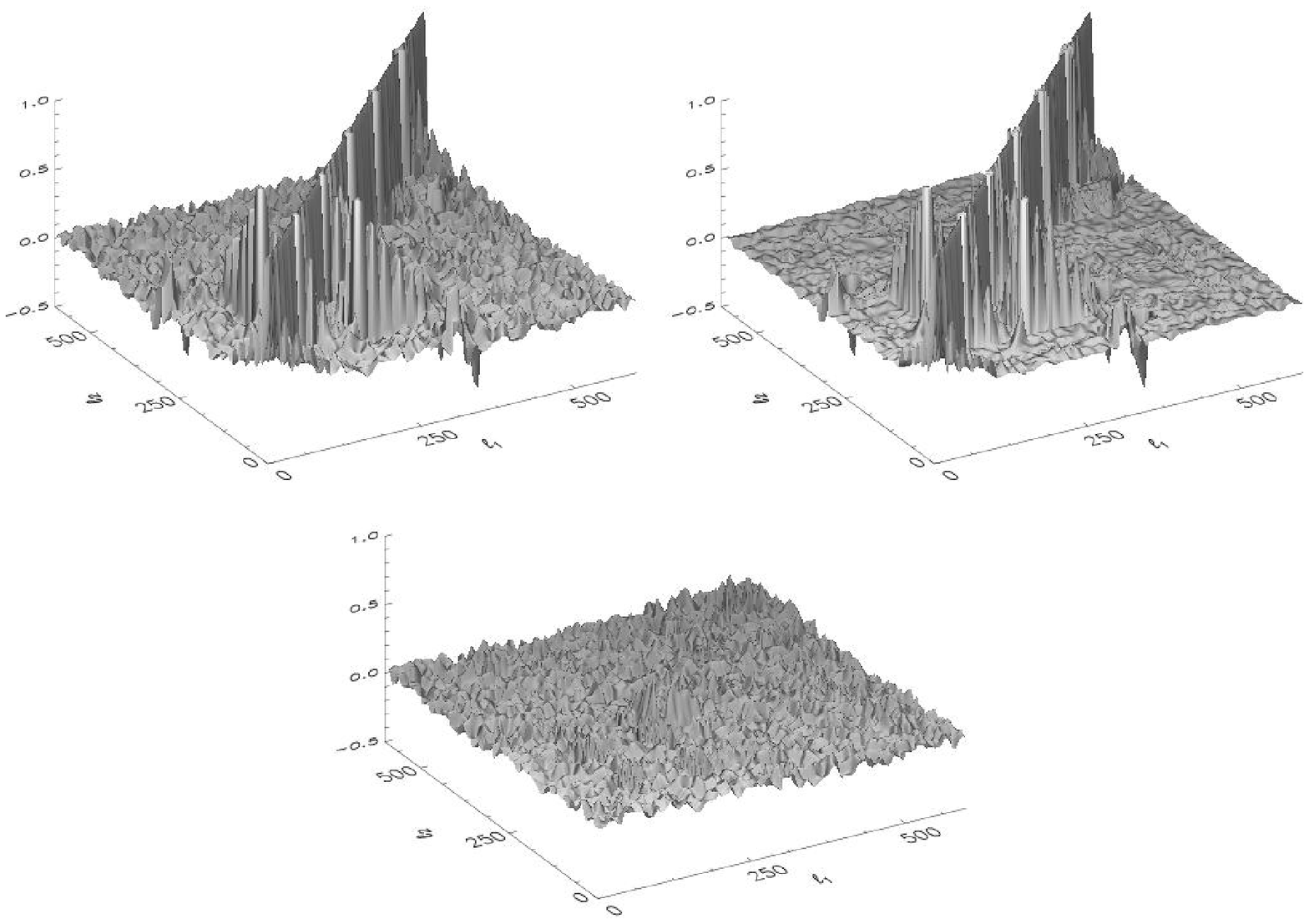}
\end{picture}
\vspace{1.5cm}
\caption{The simulated ({\it upper, left}) and semi-analytic ({\it upper,
right}) correlations for the WMAP 1$\sigma$ region QUaD simulations
shown in Figs.~\ref{cov_quad_1sigma} and \ref{diag_quad_1sigma}. The
residuals between the two, which have an \emph{rms} value of $\sim 0.04$
are shown in the lower plot. For clarity, we plot the
correlations for $1 \le \ell \le 100$ only. Multipoles in the range, 
$1\le\ell\le100$ are the TT spectra, $101\le\ell\le200$ are EE ,
$201\le\ell\le300$ are BB, $301\le\ell\le400$ are TE,
$401\le\ell\le500$ are TB and $501\le\ell\le600$ are the EB spectra. }
\label{residuals_quad}
\end{figure*}

Figs.~\ref{residuals_planck} and \ref{residuals_quad} demonstrate the
performance of the semi-analytic approximation for the off-diagonal
elements. These plots, which are for the WMAP $1\sigma$ region model 
examples shown in Figs.~\ref{cov_planck_1sigma} to
\ref{diag_quad_1sigma} are for the Planck-like
(Fig.~\ref{residuals_planck}) and QUaD-like
(Fig.~\ref{residuals_quad}) simulations respectively. In each case,
we plot the correlation matrix as estimated from the simulations and
from the semi-analytic approximation, and we also plot the residuals,
${\rm corr}(MC) - {\rm corr}(SA)$ between the two. The root-mean-squared (\emph{rms}) 
residuals between the simulated and fitted covariances are $0.033$ and 
$0.038$ for the WMAP 1$\sigma$ region Planck-like and QUaD-like
simulations respectively.
It is clear from these plots that the approximation is reproducing all 
the significant features of the pseudo-$C_\ell$ covariances. The
remaining correlations apparent in the residual plots are mostly due
to the finite number of simulations used and can be reduced to below
any desired level of significance simply by increasing the number of
MC simulations. 

Approximate formulae, such as the ``Knox'' formulae (Knox 1995;
Jungman et al. 1996) have long been used to estimate the power
spectrum error bars. In such formulae, the
fractional error on the power spectrum typically scales as
$\sqrt{2/(2\ell+1)f_{\rm sky}}$, where $f_{\rm sky}$ is the fraction of sky
observed. These formulae can also account for the effects of instrument
noise. They do not, however, take account of mode-mode coupling or 
mixing between E and B modes due to finite areas of sky. It is
interesting therefore to compare the errors predicted by such formulae
to the errors we have measured from the MC simulations and the
semi-analytic fits. We have used the expressions given in equations
(4-11) of Eisenstein, Hu \& Tegmark (1999) to calculate Knox
errors for the six power spectra. These expressions predict the errors
on the true spectra and so, we have converted them to errors on the
pseudo-$C_\ell$ spectra using 
\be
(\Delta \widetilde{\bf C}_\ell)^2 = \sum_\ld {\bf M}^2_\lld (\Delta
{\bf C}_\ld)^2,
\ee
where ${\bf M}_{\ell \ld}$ are the coupling matrices of Section
\ref{pclmethod}. The errors predicted using these formulae are
compared to the simulated and semi-analytic errors in
Figs.~\ref{diag_planck_1sigma}, \ref{diag_quad_1sigma} and 
\ref{diag_quad_2sigma} for the Planck- and QUaD-like simulations.
Although, the approximate formulae systematically underestimate the
errors over most of the $\ell$-range, for multipoles above $\ell \sim
100$, the discrepancy is more or less constant for a given
experiment. In the case of Planck, the Knox errors under-estimate the
true errors by about $10$ per cent whereas for the QUaD simulations,
the errors are under-estimated by $\sim 25$ per cent.

\vspace{-0.5cm}
\subsubsection{Implications for parameter estimation}
\label{covarsims_implications}
We have tested our approximation for the covariance matrix over a 
restricted range of cosmological parameter space and have found it
reproduces all the significant structure of the covariance measured
from simulations. 

Ultimately, one wishes to use the semi-analytic approximation for
parameter estimation and there are a number of options
available. Firstly, parameter estimation could be performed directly
on the pseudo-$C_\ell$ measurements. In this case the semi-analytic 
covariance matrix could be used as it is and the model spectra would need 
to be convolved with the coupling matrices of Section
\ref{pclmethod}. Alternatively, one could perform the estimation using
the true model spectra in which case, the pseudo-$C_\ell$ covariance
would need to be converted to full-sky $C_\ell$ covariances. For a
small-scale experiment like QUaD, one needs to reconstruct the
full-sky spectra in terms of bandpowers and so the covariance of the 
bandpowers would be needed. This is easily achieved using 
\[
\langle \Delta {\bf P}_b \, \Delta {\bf P}_{b'} \rangle = \]
\be
\hspace{1cm} {\bf M}^{-1}_{b{b_1}} \, P_{{b_1} \ell} \,  
\langle \Delta \widetilde{\mathbf C}_\ell \, \Delta \widetilde{\mathbf C}_{\ld}
\rangle \, (P_{{b_2} \ld})^T \, ({\bf M}^{-1}_{b'{b_2}})^T,   
\label{covar_bands}
\ee
where ${\bf M}_{bb'}$ and $P_{bl}$ are the binned coupling matrix and
the binning matrix respectively, defined in section 4.1.2.  

An estimate of the uncertainties in parameter estimates 
resulting from our approximation is clearly also required. We are currently 
investigating with simulations what these uncertainties are and our conclusions
regarding the effect of the covariance approximation on parameter
constraints will be presented in a future paper. However, it is clear
already that the accuracy of our approximation is a function of a
number of factors. Firstly, as
indicated by the difference in the two sets of simulations we have
conducted, the performance of the approximation improves as 
the sky coverage of the data increases. This is to be expected, of
course, since as we move closer to full-sky coverage, the
approximation converges on the theoretical covariance of the full-sky
${\bf C}_\ell$ spectra. Secondly, it is clear from the simulations that the accuracy
also depends on the cosmological models one wishes to investigate. If
the parameter estimation procedure is restricted to a certain range of
parameters (e.g. by ignoring cosmological models ruled out by previous
experiments such as WMAP, 2dFGRS and/or SDSS), then the approximation may
be used without too much loss of accuracy even for an experiment with
a large sky-cut like QUaD. We are therefore confident that our
approximation will be applicable to parameter estimation from 
forthcoming CMB temperature and polarization experiments with
differing sky survey areas such as QUaD, BICEP and/or Planck.  

%%%%%%%%%%%%%%%%%%%%%%%%%%%%%%%%%%%%%%%%%%%%%%%%%%%%%%%%%%%%%%%%%%%%%%
\vspace{-5mm}
\section{Conclusion}
Pseudo-$C_\ell$ estimators and their use in reconstructing full-sky
power spectra have become popular tools in recent years for measuring
the temperature and polarization power spectra from mega-pixel
observations of the CMB (Netterfield et al. 2002; Ruhl et al. 2002;
Beno\^it et al. 2003; Hinshaw et al. 2003). In this paper, we have investigated in 
detail the reconstruction of the full set of CMB temperature and polarization 
full-sky power spectra from the pseudo-$C_\ell$ estimators. 
We have presented the calculation of the pseudo-$C_\ell$ method
relating the observed to full-sky power and cross-power spectra for
the complete set of CMB temperature and polarization spectra:
$C^{TT}_\ell$, $C^{EE}_\ell$, $C^{BB}_\ell$, $C^{TE}_\ell$,
$C^{TB}_\ell$ and $C^{EB}_\ell$. These calculations provide one with a
mechanism to reconstruct all six CMB power spectra from observations
of the CMB over a finite region of sky. 

Having calculated the relevant coupling matrices required for the
power spectra reconstructions, we have tested these reconstructions,
which we have formulated in terms of bandpowers, on simulated CMB
temperature and polarization data sets. We have conducted these tests 
on realistic simulations of two upcoming CMB experiments -- (1) the
Planck satellite where we have considered the reconstruction of
the power spectra from two sky surveys over 14 months and (2) a
combination of the ground-based QUaD polarization experiment with the 
temperature data from 4-year WMAP observations. In both cases, we have
found the estimators to be unbiased over the entire range of
multipoles investigated between $\ell=0$ and $\ell=2500$. For these
tests, we have included primordial gravitational waves and
the effects of gravitational lensing in our input cosmological model
in order to test the reconstruction of the B-mode polarization power 
spectrum. Using an input tensor-to-scalar ratio of
$T/S=0.05$, we recover a significant detection of the lensing-induced B-mode power from
both the Planck and QUaD simulations for multipoles, $\ell \le 1000$.
The reconstructions of the non-cosmological cross-spectra, $C^{TB}_\ell$ and 
$C^{EB}_\ell$ are consistent with zero on all scales for both
experimental scenarios. The reconstruction of these two spectra can be 
useful as a test for systematic effects in either the data itself or in the
data analysis pipeline of CMB experiments. 

We have proposed a new method for fast and accurate calculation of the
covariance matrix of the pseudo-$C_\ell$ estimators for any given
cosmological model. Having presented the expressions for the full
covariance of the pseudo-$C_\ell$ estimators, we have made a
symmetrized approximation, similar to that used by Efstathiou (2004) 
for the temperature-only case. Our proposed method is based on fitting the
mixing matrices of the approximated expressions from the covariances
measured from simulated data sets. We have tested this semi-analytic
approximation by comparing it with the covariance measured from
simulations for a range of cosmological models drawn from within the
WMAP 3$\sigma$ parameter space region. These tests have been carried
out for both a Planck-like experiment, with a galactic cut applied and 
for a QUaD-like experiment covering $\sim 300$ deg$^2$. We find that
the semi-analytic approximation reproduces all the major features of
the simulated covariances with varying degrees of accuracy which
depend mainly on survey area and the range of cosmological parameters
investigated. The effect of our approximation on the estimation of
cosmological parameters will be investigated in a forthcoming
paper. Comparing the errors measured from the simulations with those
predicted by the ``Knox'' formulae, we find that the latter
under-estimate the true errors by $\sim 10$ per cent for a near
full-sky experiment and by $\sim 25$ per cent for a small-scale
experiment like QUaD which spans $\sim 0.78$ per cent of the sky. 
   
In the not-too-distant future, there is the prospect of high-quality 
full-sky mega-pixel maps of not only the CMB temperature field, but
also of the polarization field from satellite missions such as WMAP,
Planck and later perhaps, NASA's proposed Inflation Probe experiment. More
immediate prospects come from a number of novel ground-based
polarization experiments such as QUaD and BICEP, both of which are due
to begin observations in 2005. Beyond this, the CLOVER experiment in
planned for 2008. These present the astronomical 
community with the opportunity of continuing the rapid progress in
cosmology already achieved with the 2dFGRS, WMAP and SDSS programs. 
In addition to improving parameter constraints by up to a factor of 6 
(e.g. Bowden et al. 2004), such polarization information
may also provide an opportunity to detect and perhaps measure the
amount of primordial gravitational waves in the Universe
through measurements of the B-mode polarization power spectrum. Such a
measurement would constitute an important addition to our knowledge of
the physics of early Universe, possibly providing a handle on the
energy scale, and other parameters, of inflation (e.g. Kinney 1998). 
If the rapid progress in CMB observations continues as envisaged, then
the methods and results presented here can be used to quickly and
accurately extract power spectra and cosmological parameters from
CMB temperature and polarization data.

%%%%%%%%%%%%%%%%%%%%%%%%%%%%%%%%%%%%%%%%%%%%%%%%%%%%%%%%%%%%%%%%%%%%%%
\vspace{-5mm}
\section*{Acknowledgements}
We thank the QUaD team, in particular Ken Ganga for useful
discussions and the referee for useful comments. MLB acknowledges a
PPARC Postdoctoral Fellowship. PGC is supported by the PPARC through a 
Postdoctoral rolling grant. ANT thanks the PPARC for an Advanced
Fellowship. Some of the results in this paper have been derived using
the {\sevensize HEALP}ix (G\'orski, Hivon \& Wandelt 1999) and 
{\sevensize CMBFAST} (Seljak \& Zaldarriaga 1996) packages. 

%%%%%%%%%%%%%%%%%%%%%%%%%%%%%%%%%%%%%%%%%%%%%%%%%%%%%%%%%%%%%%%%%%%%%%
\vspace{-5mm}
\section*{References}

 \bib Beno\'it A. et al., 2003, A\&A, 399, L19

 \bib Bennet C. et al., 2003, ApJS, 148, 1

 \bib Bond J. R., Jaffe A. H., Knox L., 1998, Phys. Rev. D., 57, 2117

 \bib Bowden M. et al., 2004, MNRAS, 349, 321
 
 \bib Bunn, E. F., 1995, PhD thesis, University of California, Berkeley

 \bib Chon G., Challinor A., Prunet S., Hivon E., Sazpudi I., 2004,
 MNRAS, 350, 914 

 \bib Challinor A., Chon G., 2004, submitted to MNRAS, astro-ph/0410097

 \bib Church S. et al., 2003, New Astron. Rev., 47, 1083

 \bib Dodelson S., Kinney W. H., Kolb E. W., 1997, Phys. Rev. D, 56,
 3207-15

 \bib Dor\'e O., Knox L., Peel A., 2001, Phys. Rev. D, 64, 3001

 \bib Eisenstein D. E., Hu W., Tegmark M., 1999, ApJ, 518, 2 

 \bib Efstathiou G., 2004, MNRAS, 349, 603

 \bib G\'orski K. M., 1994, ApJ, 430, L85

 \bib G\'orski K. M. et al., 1994, ApJ, 430, L89

 \bib G\'orski K. M., Hivon E., Wandelt B. D., 1999, in Banday A. J.,
 Sheth R. S., Da Costa L., eds., Proceedings of the MPA/ESO Cosmology
 Conference ``{\it Evolution of Large-Scale Structure}'', PrintPartners
 Ipskamp, NL, pp. 37-42

 \bib Hansen F., G\'orski K. M., 2003, MNRAS, 343, 559

 \bib Hansen F., G\'orski K. M., Hivon E., 2002, MNRAS, 336, 1304

 \bib Hinshaw G. et al., 2003, ApJS, 148, 135

 \bib Hivon E., G\'orski K. M., Netterfield C. B., Crill B. P., Prunet
 S., Hansen F., 2002, ApJ, 567, 2

 \bib Jewell J., Levin S., Anderson C. H., 2004, ApJ, 609, 1

 \bib Jungman G., Kamionkowski M., Kosowsky A., Spergel D. N., 1996,
 Phys. Rev. D, 54, 1332

 \bib Kamionkowski M., Kosowsky A., Stebbins A., 1997,
 Phys. Rev. Lett., 78, 2058

 \bib Kamionkowski M., Kosowsky A., 1998, Phys. Rev. D, 57, 685

 \bib Kinney W. H., 1998, Phys. Rev. D, 58, 123506

 \bib Knox L., 1995, Phys. Rev. D, 52, 4307

 \bib Kogut A. et al., 2003, ApJS, 148, 161

% \bib Kovac J. M., Leitch E. M., Pryke C., Carlstrom J. E., Halverson
% N. W., Holzapfel W. L., 2002, Nature, 420, 772
 \bib Kovac J. M. et al., 2002, Nature, 420, 772

 \bib Lepora N., 1998, preprint, gr-qc/9812077

 \bib Lewis A., Challinor A., Turok N., 2002, Phys. Rev. D, 65, 023505

 \bib Liddle A. R., Lyth D. H., 2000, {\it Cosmological Inflation and
 Large-Scale Structure}, Cambridge University Press, Cambridge. 

 \bib Lue A., Wang L., Kamionkowski M., 1999, Phys. Rev. Lett., 83,
 1506 

 \bib Netterfield C. B. et al., 2002, ApJ, 571, 604

 \bib Peebles P. J. E., 1973, ApJ, 185, 413

 \bib Percival W. J. et al., 2001, 327, 1297

 \bib Readhead A. C. S. et al., 2004, Science (published online
 October 7, 2004), astro-ph/0409569

 \bib Ruhl J. E. et al., 2003, ApJ, 599, 786

 \bib Schulten K., Gordon R. G., 1975, J. Math. Phys., 16, 1961

 \bib Seljak U., Zaldarriaga M., 1996, ApJ, 469, 437

 \bib Seljak U., Zaldarriaga M., 1997, Phys. Rev. Lett., 78, 2054

 \bib Spergel D. N. et al., 2003, ApJS, 148, 175

 \bib Szapudi I., Prunet S., Colombi S., 2001, ApJ, 561, L11

 \bib Szapudi I., Prunet S., Pogosyan D., Szalay A. S., Bond J. R.,
 2001, ApJ, 548, L115

 \bib Tauber J. A., 2004, Advances in Space Research, 34, 491

 \bib Taylor A. C. et al., 2004, to appear in the proceedings of the
 XXXVIXth Rencontres de Moriond ``{\it Exploring the Universe}", astro-ph/0407148 

 \bib Tegmark M., 1997, Phys. Rev. D, 55, 5895

 \bib Tegmark M. et al., 2004, ApJ, 606, 702    

 \bib Tristram M., Macias-Perez J. F., Renault C., Santos C., 2005,
 MNRAS, in press, astro-ph/0405575

 \bib Varshalovich D. A., Moskalev A. N., Khersonskii V. K., 1998,
 {\it Quantum Theory of Angular Momentum}, World Scientific, Singapore

 \bib Verde L. et al., 2003, ApJS, 148, 195

 \bib Wandelt B. D., Hivon E., G\'orski K. M., 2001, Phys. Rev. D, 64,
 083003

 \bib Wandelt B. D., Larson D. L., Lakshminarayanan A., 2004,
 preprint, astro-ph/0310080

 \bib Zaldarriaga M., Seljak U., 1997, Phys. Rev. D, 55, 1830

%%%%%%%%%%%%%%%%%%%%%%%%%%%%%%%%%%%%%%%%%%%%%%%%%%%%%%%%%%%%%%%%%%%%%%

\onecolumn
\appendix

%%%%%%%%%%%%%%%%%%%%%%%%%%%%%%%%%%%%%%%%%%%%%%%%%%%%%%%%%%%%%%%%%%%%%%

\section{Pseudo-$C_\ell$ coupling matrices}
\label{appendixA}
As discussed in Section \ref{stats_cutsky}, the pseudo-$C_\ell$
spectra measured from a finite region of sky are related to the
underlying full-sky power spectra by the matrix relation, 
\be
\widetilde{\bf C} = \sum_\ld {\bf M}_\lld {\bf C}_\ld,
\ee
which can be expanded as
\be 
\left( \begin{array}{c} 
\tilde{C}^{TT}_{\ell} \\ \tilde{C}^{TE}_{\ell} \\
\tilde{C}^{TB}_{\ell} \\ \tilde{C}^{EE}_{\ell} \\
\tilde{C}^{EB}_{\ell} \\ \tilde{C}^{BB}_{\ell} \end{array} \right) 
= \sum_{\ell'} \left( \begin{array}{cccccc} 
W^{00}_\lld & 0 & 0 & 0 &  0 & 0 \\ 
0 & W^{0+}_\lld & W^{0-}_\lld & 0 & 0 & 0 \\ 
0 &-W^{0-}_\lld & W^{0+}_\lld & 0 & 0 & 0 \\ 
0 & 0 & 0 & W^{++}_\lld & (W^{-+}_\lld+W^{+-}_\lld) & W^{--}_\lld \\ 
0 & 0 & 0 & -W^{+-}_\lld& (W^{++}_\lld-W^{--}_\lld) & W^{-+}_\lld \\ 
0 & 0 & 0 & W^{--}_\lld & -(W^{-+}_\lld+W^{+-}_\lld)& W^{++}_\lld \\ 
\end{array} \right) 
\left( \begin{array}{c} 
C^{TT}_{\ell'} \\ C^{TE}_{\ell'} \\ C^{TB}_{\ell'} \\ 
C^{EE}_{\ell'} \\ C^{EB}_{\ell'} \\ C^{BB}_{\ell'} \end{array} \right) 
\label{eq:pseudo_full_expression},
\ee 
where we have defined
\be W^{MN}_\lld = \frac{1}{2 \ell +1} 
\sum_{mm'}\,
  W^M_{\ell\ell'mm'}\, (W^N_{\ell'\ell m'm})^* 
\label{eq:mixing_matrices}
\ee 
for $M,N=(0,+,-)$. The $W^{0/+/-}_{\ell\ell'mm'}$ matrices are defined 
as in equations~(\ref{eq:w_1}), (\ref{eq:w_2}) and (\ref{eq:w_3}). 
In order to relate ${\mathbf {\widetilde C}}_{\ell}$ to ${\mathbf C}_{\ld}$,
one needs to numerically calculate expressions of the type 
\be
\sum_{mm'}\, _s W^\mmd_\lld (_{s'} W^{m'm}_{\ell'\ell})^*,
\label{eq:sum_ww}
\ee
with $_s W^\mmd_\lld$ given by equation~(\ref{eq:w_1}) or
(\ref{eq:w_3}). These are cumbersome to evaluate due to the high
number of summations over multipoles aggravated by the need to perform 
integrals over the whole sphere. Fortunately, the coupling matrix, 
${\mathbf M}_{\ell \ld}$, can be simplified using the properties of the 
Wigner-3j symbols. One starts by expanding the window functions, 
$W_T(\Omega)$ and $W_P(\Omega)$, in spin-0 spherical harmonics such
that their auto- and cross-power spectra (which we have scaled here by
$2\ell+1$) are given by
\ba
{\mathcal W}^{TT}_{\ell} = \sum_{m}w^T_{\lm}(w^T_{\lm})^*, \nn 
{\mathcal W}^{PP}_{\ell} = \sum_{m}w^P_{\lm}(w^P_{\lm})^*, \nn
{\mathcal W}^{TP}_{\ell} = \sum_{m}w^T_{\lm}(w^P_{\lm})^*, 
\label{eq:window_power}
\ea 
where the $w^T_\lm$ and $w^P_\lm$ coefficients are the spherical
harmonic transform coefficients of the window functions:
\ba
w^T_\lm = \int d \Omega \, W_T(\Omega) Y^*_\lm, \nn 
w^P_\lm = \int d \Omega \, W_P(\Omega) Y^*_\lm. 
\ea
In order to calculate terms like equation~(\ref{eq:sum_ww}),
we apply the properties of integrals over solid
angle of the spin-weighted spherical harmonics, 
\ba
\int d \Omega \,_sY_\lm^* (\Omega) \,_{s'}Y_\lmd (\Omega)
\,_{s''}Y_\lmdd (\Omega) =(-1)^{m+s} {\sqrt
\frac{(2\ell+1)(2\ld+1)(2\ldd+1)}{4\pi} }\nonumber\\ 
\left( \begin{array}{ccc} \ell & \ld & \ldd \\ 
 s & -s' & -s'' \end{array} \right) 
\left( \begin{array}{ccc} \ell & \ld & \ell ''' \\ 
-m & m' & m''' \end{array} \right), 
\ea 
\vspace{2mm} 
and finally use the orthogonality relation of the Wigner 3-j symbols
(Varshalovich, Moskalev \& Khersonskii 1998), 
\be 
(2\ell''+1) \sum_{mm'}
\left( \begin{array}{ccc} \ell & \ld & \ldd \\ 
m & m' & m'' \end{array} \right) 
\left( \begin{array}{ccc} \ell & \ld & \ell''' \\ 
m & m' & m''' \end{array} \right) 
= \delta_{\ldd \ell'''}\delta_{m''m'''}.  
\ee
Then it is straightforward to notice that all the imaginary terms 
present in the coupling matrix, ${\mathbf M}_{\ell \ld}$, vanish because 
of the symmetries of the Wigner 3-j symbols, in particular 
\be 
\left( \begin{array}{ccc} \ell & \ld & \ldd \\ 
m & m' & m'' \end{array} \right) 
= (-1)^{\ell+\ld+\ldd} \left( \begin{array}{ccc} \ell & \ld & \ldd \\ 
-m & -m' & -m'' \end{array} \right).  
\ee 
The imaginary terms are either associated with the
expressions $W^{0-}_\lld$ or $W^{+-}_\lld$ (and their complex
conjugates) in equation~(\ref{eq:pseudo_full_expression}). 
For the term $W^{0-}_\lld$ , we have a dependency of the type, 
\ba
F(\ell,\ld,\ldd) &=& \left( \begin{array}{ccc} \ell & \ld & \ldd \\ 
0 & 0 & 0 \end{array} \right) 
\left[ \left( \begin{array}{ccc} \ell & \ld & \ldd \\ 
-2 & 2 & 0 \end{array} \right)
- \left( \begin{array}{ccc} \ell & \ld & \ldd \\ 
2 & -2 & 0 \end{array} \right) \right] \nonumber\\ 
&=& (-1)^{\ell+\ld+\ldd} 
\left( \begin{array}{ccc} \ell & \ld & \ldd \\ 
0 & 0 & 0 \end{array} \right) 
\left[ \left( \begin{array}{ccc} \ell & \ld & \ldd \\ 
2 & -2 & 0 \end{array} \right)
- \left( \begin{array}{ccc} \ell & \ld & \ldd \\ 
-2 & 2 & 0 \end{array} \right) \right] \nonumber \\ 
&=& \left( \begin{array}{ccc} \ell & \ld & \ldd \\ 
0 & 0 & 0 \end{array} \right) 
\left[ \left( \begin{array}{ccc} \ell & \ld & \ldd \\ 
2 & -2 & 0 \end{array} \right)
- \left( \begin{array}{ccc} \ell & \ld & \ldd \\ 
-2 & 2 & 0 \end{array} \right) \right] \nonumber\\ 
&=& -F(\ell,\ld,\ldd), 
\ea 
thus $F(\ell,\ld,\ldd)=0$. The same line of reasoning applies for the 
term $W^{+-}_\lld$. 
\vspace{2mm} 
Therefore, we obtain the simplified relation, 
\be 
\sum_{mm'}\, _s W^\mmd_\lld (_{s'} W^{m'm}_{\ell'\ell})^* = 
\frac{(2\ell+1)(2\ld+1)}{4\pi}\sum_\ldd {\mathcal W}_\ldd^{SS'} 
\left( \begin{array}{ccc} \ell & \ld & \ldd \\
-s & s & 0 \end{array} \right) 
\left( \begin{array}{ccc} \ell & \ld & \ldd \\ 
-s' & s' & 0 \end{array} \right),
\ee 
where $S = T$ if $s=0$  and $S=P$ if $s=+2$ or $-2$. The value of $S'$
depends in the same way on the value of $s'$.  
Using the previous result and after some algebra, we obtain for the
non-zero mode-mode coupling matrix ${\mathbf M}_{\ell \ld}$ components 
(see equation~\ref{pcl1}) the following set of expressions:
\ba
M_{\ell\ld}^{TT,TT} & = & \frac{(2\ld+1)}{4\pi}
\sum_{\ldd}{\mathcal W}_{\ldd}^{TT} 
\left( \begin{array}{ccc} \ell & \ld & \ldd \\ 
0 & 0 & 0 \end{array} \right)^2 \label{eq: M_TT} \\
M_{\ell\ld}^{TE,TE} & = & M_{\ell\ld}^{TB,TB}  \nonumber\\
& = &\frac{(2\ld+1)}{4\pi} \sum_{\ldd}{\mathcal W}_{\ldd}^{TP} 
\left( \begin{array}{ccc} \ell & \ld & \ldd \\ 
0 & 0 & 0 \end{array} \right) 
\left( \begin{array}{ccc} \ell & \ld & \ldd \\ 
2 & -2 & 0 \end{array} \right) \\
M_{\ell\ld}^{EE,EE}& = & M_{\ell\ld}^{BB,BB} \nonumber\\ 
& =  &\frac{(2\ld+1)}{8\pi} \sum_{\ldd}{\mathcal W}_{\ldd}^{PP}
\left[1 + (-1)^{\ell+\ld+\ldd} \right]
\left( \begin{array}{ccc} \ell & \ld & \ldd \\ 
2 & -2 & 0 \end{array} \right)^2 \\ 
M_{\ell\ld}^{EE,BB} & = & M_{\ell\ld}^{BB,EE} \nonumber\\
& = &\frac{(2\ld+1)}{8\pi} \sum_{\ldd}{\mathcal W}_{\ldd}^{PP} 
\left[ (-1)^{\ell+\ld+\ldd} -1 \right]
\left( \begin{array}{ccc} \ell & \ld & \ldd \\ 
2 & -2 & 0 \end{array} \right)^2 \\
M_{\ell\ld}^{EB,EB} & = & \frac{(2\ld+1)}{4\pi}
\sum_{\ldd}{\mathcal W}_{\ldd}^{PP} 
\left( \begin{array}{ccc} \ell & \ld & \ldd \\ 
2 & -2 & 0 \end{array} \right)^2 \\
\label{eq: M_EB} 
\ea
%%%%%%%%%%%%%%%%%%%%%%%%%%%%%%%%%%%%%%%%%%%%%%%%%%%%%%%%%%%%%%%%%%%%%%

\section{Covariance matrix of the pseudo-$C_\ell$ estimators}
\label{appendixB}
In Section~\ref{pclcovar}, we have described our procedure for
calculating the semi-analytic approximation for the pseudo-$C_\ell$
covariance matrix. Here, we present the explicit expressions for the
semi-analytic approximation. We write these in terms of the various 
mixing matrices, ${\bf X}^{abcd}_{\ell\ld}$, which we define by 
\be
{\bf X}^{abcd}_{\lld} = \frac{1}{(2\ell+1)(2\ld+1)}
\sum_{mm'} \sum_{\lone \mone} \sum_{\ltwo \mtwo}
W^{\,\,\,a}_{\ell\lone m\mone} W^{* \,\, b}_{\ld \lone m'\mone}  
W^{\,\,\,c}_{\ld\ltwo m'\mtwo} W^{* \,\, d}_{\ell \ltwo m\mtwo},  
\ee
where $\{a,b,c,d\}$ can take any of the values $\{0,+,-\}$ and the $W$
matrices are the entries of the window matrix, 
equation~(\ref{eq:window_matrix}), and are defined in
equations~(\ref{eq:w_1}), (\ref{eq:w_2}) and (\ref{eq:w_3}). In terms of these matrices
then, the approximations to the various pseudo-$C_\ell$ covariances
are:

\be
\lgl \Delta \widetilde{C}^{TT}_\ell \Delta \widetilde{C}^{TT}_\ld \rgl
\approx 2 \, C^{TT}_\ell C^{TT}_\ld \, {\bf X}^{0000}_\lld,
\label{eq:app2}
\ee

\be
\lgl \Delta \widetilde{C}^{TE}_\ell \Delta \widetilde{C}^{TE}_\ld \rgl
\approx C^{TE}_\ell C^{TE}_\ld \, {\bf X}^{0+0+}_\lld
+ \sqrt{C^{TT}_\ell C^{TT}_\ld C^{EE}_\ell C^{EE}_\ld} \, {\bf X}^{00++}_\lld
+ \sqrt{C^{TT}_\ell C^{TT}_\ld C^{BB}_\ell C^{BB}_\ld} \, {\bf X}^{00--}_\lld,
\label{eq:app3}
\ee

\be
\lgl \Delta \widetilde{C}^{TB}_\ell \Delta \widetilde{C}^{TB}_\ld \rgl
\approx C^{TE}_\ell C^{TE}_\ld \, {\bf X}^{0-0-}_\lld
+ \sqrt{C^{TT}_\ell C^{TT}_\ld C^{EE}_\ell C^{EE}_\ld} \, {\bf X}^{00--}_\lld
+ \sqrt{C^{TT}_\ell C^{TT}_\ld C^{BB}_\ell C^{BB}_\ld} \, {\bf X}^{00++}_\lld,
\ee

\be
\lgl \Delta \widetilde{C}^{EE}_\ell \Delta \widetilde{C}^{EE}_\ld \rgl
\approx 2 \, C^{EE}_\ell C^{EE}_\ld \, {\bf X}^{++++}_\lld
+       2 \, C^{BB}_\ell C^{BB}_\ld \, {\bf X}^{----}_\lld
+ 4\sqrt{C^{EE}_\ell C^{EE}_\ld C^{BB}_\ell C^{BB}_\ld} \, {\bf X}^{++--}_\lld, 
\ee

\be
\lgl \Delta \widetilde{C}^{BB}_\ell \Delta \widetilde{C}^{BB}_\ld \rgl
\approx 2 \, C^{BB}_\ell C^{BB}_\ld \, {\bf X}^{++++}_\lld
+       2 \, C^{EE}_\ell C^{EE}_\ld \, {\bf X}^{----}_\lld
+ 4\sqrt{C^{EE}_\ell C^{EE}_\ld C^{BB}_\ell C^{BB}_\ld} \, {\bf X}^{++--}_\lld,
\ee

\vspace{-5mm}
\ba
\lgl \Delta \widetilde{C}^{EB}_\ell \Delta \widetilde{C}^{EB}_\ld \rgl
&\approx& 
C^{EE}_\ell C^{EE}_\ld \, \left[ {\bf X}^{+-+-}_\lld + {\bf X}^{++--}_\lld \right]
+ C^{BB}_\ell C^{BB}_\ld \, \left[ {\bf X}^{-+-+}_\lld + {\bf
X}^{--++}_\lld \right] \nn
&+& \sqrt{C^{EE}_\ell C^{EE}_\ld C^{BB}_\ell C^{BB}_\ld} \, 
\left[ {\bf X}^{++++}_\lld + {\bf X}^{----}_\lld - {\bf X}^{+--+}_\lld -
{\bf X}^{-++-}_\lld \right],
\ea

\be
\lgl \Delta \widetilde{C}^{TT}_\ell \Delta \widetilde{C}^{TE}_\ld \rgl
\approx 2 \, \sqrt{C^{TT}_\ell C^{TT}_\ld C^{TE}_\ell C^{TE}_\ld} \, {\bf X}^{0+00}_\lld,
\ee

\be
\lgl \Delta \widetilde{C}^{TT}_\ell \Delta \widetilde{C}^{TB}_\ld \rgl
\approx -2 \, \sqrt{C^{TT}_\ell C^{TT}_\ld C^{TE}_\ell C^{TE}_\ld} \, {\bf X}^{00-0}_\lld,
\ee

\be
\lgl \Delta \widetilde{C}^{TT}_\ell \Delta \widetilde{C}^{EE}_\ld \rgl
\approx 2 \, C^{TE}_\ell C^{TE}_\ld \, {\bf X}^{0++0}_\lld,
\ee

\be
\lgl \Delta \widetilde{C}^{TT}_\ell \Delta \widetilde{C}^{BB}_\ld \rgl
\approx 2 \, C^{TE}_\ell C^{TE}_\ld \, {\bf X}^{0--0}_\lld,
\ee

\be
\lgl \Delta \widetilde{C}^{TT}_\ell \Delta \widetilde{C}^{EB}_\ld \rgl
\approx -2 \, C^{TE}_\ell C^{TE}_\ld \, {\bf X}^{0+-0}_\lld,
\ee

\be
\lgl \Delta \widetilde{C}^{TE}_\ell \Delta \widetilde{C}^{TB}_\ld \rgl
\approx - \, C^{TE}_\ell C^{TE}_\ld \, {\bf X}^{0-0+}_\lld
- \sqrt{C^{TT}_\ell C^{TT}_\ld C^{EE}_\ell C^{EE}_\ld} \, {\bf X}^{00-+}_\lld
+ \sqrt{C^{TT}_\ell C^{TT}_\ld C^{BB}_\ell C^{BB}_\ld} \, {\bf X}^{00+-}_\lld,
\ee

\be
\lgl \Delta \widetilde{C}^{TE}_\ell \Delta \widetilde{C}^{EE}_\ld \rgl
\approx 2 \, \sqrt{C^{TE}_\ell C^{TE}_\ld C^{EE}_\ell C^{EE}_\ld} \, {\bf X}^{0+++}_\lld
+ 2 \sqrt{C^{TE}_\ell C^{TE}_\ld C^{BB}_\ell C^{BB}_\ld} \, {\bf X}^{0+--}_\lld,
\ee

\be
\lgl \Delta \widetilde{C}^{TE}_\ell \Delta \widetilde{C}^{BB}_\ld \rgl
\approx 2 \, \sqrt{C^{TE}_\ell C^{TE}_\ld C^{EE}_\ell C^{EE}_\ld} \, {\bf X}^{0--+}_\lld
- 2 \sqrt{C^{TE}_\ell C^{TE}_\ld C^{BB}_\ell C^{BB}_\ld} \, {\bf X}^{0-+-}_\lld,
\ee

\be
\lgl \Delta \widetilde{C}^{TE}_\ell \Delta \widetilde{C}^{EB}_\ld \rgl
\approx \sqrt{C^{TE}_\ell C^{TE}_\ld C^{BB}_\ell C^{BB}_\ld} \, 
\left[ {\bf X}^{0++-}_\lld - {\bf X}^{0---}_\lld \right]
- \sqrt{C^{TE}_\ell C^{TE}_\ld C^{EE}_\ell C^{EE}_\ld} \, 
\left[ {\bf X}^{0-++}_\lld + {\bf X}^{0+-+}_\lld \right],
\ee

\be
\lgl \Delta \widetilde{C}^{TB}_\ell \Delta \widetilde{C}^{EE}_\ld \rgl
\approx 2 \, \sqrt{C^{TE}_\ell C^{TE}_\ld C^{BB}_\ell C^{BB}_\ld} \, {\bf X}^{0+-+}_\lld
- 2 \sqrt{C^{TE}_\ell C^{TE}_\ld C^{EE}_\ell C^{EE}_\ld} \, {\bf X}^{0++-}_\lld,
\ee

\be
\lgl \Delta \widetilde{C}^{TB}_\ell \Delta \widetilde{C}^{BB}_\ld \rgl
\approx -2 \sqrt{C^{TE}_\ell C^{TE}_\ld C^{EE}_\ell C^{EE}_\ld} \, {\bf X}^{0---}_\lld
- 2 \sqrt{C^{TE}_\ell C^{TE}_\ld C^{BB}_\ell C^{BB}_\ld} \, {\bf X}^{0-++}_\lld,
\ee

\be
\lgl \Delta \widetilde{C}^{TB}_\ell \Delta \widetilde{C}^{EB}_\ld \rgl
\approx \sqrt{C^{TE}_\ell C^{TE}_\ld C^{EE}_\ell C^{EE}_\ld} \, 
\left[ {\bf X}^{0-+-}_\lld - {\bf X}^{0+--}_\lld \right]
+ \sqrt{C^{TE}_\ell C^{TE}_\ld C^{BB}_\ell C^{BB}_\ld} \, 
\left[ {\bf X}^{0+++}_\lld - {\bf X}^{0--+}_\lld \right],
\ee

\be
\lgl \Delta \widetilde{C}^{EE}_\ell \Delta \widetilde{C}^{BB}_\ld \rgl
\approx 2 \, C^{EE}_\ell C^{EE}_\ld \, {\bf X}^{+--+}_\lld
+       2 \, C^{BB}_\ell C^{BB}_\ld \, {\bf X}^{-++-}_\lld
- 4\sqrt{C^{EE}_\ell C^{EE}_\ld C^{BB}_\ell C^{BB}_\ld} \, {\bf X}^{+-+-}_\lld,
\ee

\vspace{-5mm}
\ba
\lgl \Delta \widetilde{C}^{EE}_\ell \Delta \widetilde{C}^{EB}_\ld \rgl
&\approx& 
2 \, C^{BB}_\ell C^{BB}_\ld \, {\bf X}^{--+-}_\lld
- 2 \, C^{EE}_\ell C^{EE}_\ld \, {\bf X}^{+-++}_\lld  
+ 2 \, \sqrt{C^{EE}_\ell C^{EE}_\ld C^{BB}_\ell C^{BB}_\ld} \,
\left[ {\bf X}^{+++-}_\lld - {\bf X}^{+---}_\lld \right],
\ea

\vspace{-5mm}
\ba
\lgl \Delta \widetilde{C}^{BB}_\ell \Delta \widetilde{C}^{EB}_\ld \rgl
&\approx& 
2 \, C^{BB}_\ell C^{BB}_\ld \, {\bf X}^{++-+}_\lld 
- 2 \, C^{EE}_\ell C^{EE}_\ld \, {\bf X}^{--+-}_\lld  
+ 2 \, \sqrt{C^{EE}_\ell C^{EE}_\ld C^{BB}_\ell C^{BB}_\ld} \, 
\left[ {\bf X}^{---+}_\lld - {\bf X}^{-+++}_\lld \right].
\label{eq:app22}
\ea

%%%%%%%%%%%%%%%%%%%%%%%%%%%%%%%%%%%%%%%%%%%%%%%%%%%%%%%%%%%%%%%%%%%%%%

\bsp
\label{lastpage}
\end{document}